\documentclass[fleqn,usenatbib,useAMS]{mnras}

\usepackage{graphicx}	
\usepackage{amsmath}	
\usepackage{amssymb}	
\usepackage{multicol}        
\usepackage{bm}		
\usepackage{pdflscape}	
\usepackage{epstopdf}
\usepackage[T1]{fontenc}
\usepackage{ae,aecompl}
\usepackage[dvipsnames]{xcolor}

\newcommand{\en}[1]{\textcolor{black}{#1}}
\newcommand{\rev}[1]{\textcolor{black}{#1}}

\newcommand{\beq}{\begin{equation}}
\newcommand{\beqa}{\begin{eqnarray}}
\newcommand{\eeq}{\end{equation}}
\newcommand{\eeqa}{\end{eqnarray}}

\newcommand{\simgt}{\lower.5ex\hbox{$\; \buildrel > \over \sim \;$}}
\newcommand{\simlt}{\lower.5ex\hbox{$\; \buildrel < \over \sim \;$}}

\newcommand{\bd}[1]{\mbox{\boldmath $#1$}}

\newcommand{\bfred}[1]{\textcolor{red}{\bf{#1}}}

\newcommand{\bx}{{\bf x}}

\newcommand{\bk}{{\bf k}}
\newcommand{\bmtheta}{{\bm{\theta}}}
\newcommand{\bmell}{{\bm{\ell}}}

\usepackage[T1]{fontenc}
\usepackage{ae,aecompl}

\usepackage{newtxtext,newtxmath}

\title[Halo-DM cross-correlations]{
Probing cosmology and gastrophysics with fast radio bursts:~\\
Cross-correlations of dark matter haloes and cosmic dispersion measures
}

\author[M. Shirasaki et al.]
{
Masato Shirasaki$^{1,2}$\thanks{Contact e-mail: \href{mailto:masato.shirasaki@nao.ac.jp}{masato.shirasaki@nao.ac.jp}},
Ryuichi Takahashi$^{3}$,
Ken Osato$^{4,5}$,
and Kunihito Ioka$^{4}$
\\
$^{1}$National Astronomical Observatory of Japan, Mitaka, Tokyo 181-8588, Japan \\
$^{2}$The Institute of Statistical Mathematics, Tachikawa, Tokyo 190-8562, Japan \\
$^{3}$Faculty of Science and Technology, Hirosaki University, Hirosaki, Aomori 036-8561, Japan \\
$^{4}$Center for Gravitational Physics, Yukawa Institute for Theoretical Physics, Kyoto University, Kyoto 606-8502, Japan\\
$^{5}$LPENS, D\'{e}partment de Physique, \'{E}cole Normale Sup\'{e}rieure, Universit\'{e} PSL, CNRS, Sorbonne Universit\'{e}, Universit\'{e} de Paris, 75005 Paris, France
}

\pubyear{2021}

\begin{document}
\label{firstpage}
\pagerange{\pageref{firstpage}--\pageref{lastpage}}
\maketitle

\begin{abstract}
For future surveys of fast radio bursts (FRBs),
we clarify information available from cosmic dispersion measures (DMs) through cross-correlation analyses
of foreground dark matter haloes (hosting galaxies and galaxy clusters) with their known redshifts.
With a halo-model approach, we predict
that the cross-correlation with cluster-sized haloes is less affected by the details of gastrophysics, providing robust cosmological information.
For less massive haloes, the cross-correlation at angular scales of $<10\, \mathrm{arcmin}$ is sensitive to gas expelled from the halo centre due to galactic feedback.
Assuming $20000$ FRBs over $20000 \, {\rm deg}^2$ with a localisation error being 3 arcmin, we expect that the cross-correlation signal at halo masses of $10^{12}$--$10^{14}\, M_\odot$
can be measured with a level of $\sim 1\%$ precision in a redshift range of $0<z<1$.
Such precise measurements enable one to put a 
1.5\% level constraint on $\sigma_8\, (\Omega_\mathrm{M}/0.3)^{0.5}$ 
and a 3\% level constraint on $(\Omega_\mathrm{b}/0.049)(h/0.67)(f_\mathrm{e}/0.95)$
($\sigma_8$, $\Omega_\mathrm{M}$, $\Omega_\mathrm{b}$, $h$ and $f_\mathrm{e}$
are the linear mass variance smoothed at $8\, h^{-1}\mathrm{Mpc}$, mean mass density, mean baryon density, the present-day Hubble parameter and fraction of free electrons in cosmic baryons today), whereas the gas-to-halo mass relation in galaxies and clusters can be constrained with a level of $10\%$--$20\%$.
Furthermore the cross-correlation analyses can break the degeneracy among 
$\Omega_\mathrm{b}$, $h$ and $f_\mathrm{e}$, inherent in the DM-redshift relation.
\rev{Our proposal opens new possibilities for FRB cosmology, while it requires extensive galaxy redshift catalogues and further improvement of the halo model.}
\end{abstract}





\section{Introduction}

Understanding the evolution of cosmic baryons is crucial 
in modern astronomy.
The mean energy density of cosmic baryons has been tightly constrained from measurements of temperature and polarisation fluctuations in the cosmic microwave background \citep[CMB; e.g.][]{2013ApJS..208...20B, 2020A&A...641A...6P} and primordial deuterium abundance based on big-bang
nucleosynthesis \citep{2018ApJ...855..102C}.
Although these probes have provided 
information about cosmic baryons in the early universe \citep[e.g.][]{1970ApJ...162..815P, 1970Ap&SS...7....3S}, late-time information \en{is required} for a complete understanding of cosmic baryons.
The cosmic baryons form large-scale structures through gravitational collapse with dark matter,
whereas their cooling and heating processes play an essential role in forming stars and galaxies.
Observational estimates have shown that the star formation in galaxies is an inefficient process, even at present, and a large fraction of cosmic baryons would remain in the gaseous phase of the universe \citep[e.g.][]{1998ApJ...503..518F, 2004ApJ...616..643F}.
This implies that most baryonic matter in the late-time universe has not been observed, thereby motivating us to develop observational methods to seek the so-called missing baryons.

Most missing baryons are expected to be found in diffuse intergalactic media,
which are too faint to detect on an individual basis \citep[see e.g.][for recent observational efforts]{2016MNRAS.457.4236B, 2018Natur.558..406N}.
Various observational probes have been proposed to study diffuse baryonic matter between galaxies and stars \citep[e.g.][for reviews]{2007ARA&A..45..221B, 2016ARA&A..54..313M, 2020ARA&A..58..363P}.
Among them, we study the dispersion measure (DM) defined as the column density of free electrons in this paper.

The DM is defined as the integral of the number density of free electrons along a line-of-sight direction.
The DM has been commonly measured from a frequency-dependent arrival time in radio pulses, allowing us to study electrons in the vicinity of the Milky Way with pulsars so far \citep[e.g.][]{1993ApJ...411..674T, 2001ApJ...553..367C, 2002astro.ph..7156C, 2017ApJ...835...29Y}.
New milli-second radio transients located at extragalactic distances, referred to as fast radio bursts \citep[FRBs;][for a recent review]{2019ARA&A..57..417C}, have opened a new window for studying the statistical properties of free electrons in an expanding universe
\citep[e.g.][]{2003ApJ...598L..79I, 2004MNRAS.348..999I, 2020Natur.581..391M}.
Ongoing radio transient surveys are aimed at constructing a large sample of FRBs and revealing their origin.
They include the Australian Square Kilometre Array Pathfinder (ASKAP\footnote{\url{https://www.atnf.csiro.au/projects/askap/index.html}}) and the Canadian Hydrogen Intensity Mapping Experiment (CHIME\footnote{\url{https://chime-experiment.ca/en}}), already providing information about hundreds of FRBs over a sky \citep[e.g.][]{2017arXiv171008155P, 2021arXiv210604352T}.
\rev{Future} surveys, such as the Square Kilometre Array (SKA\footnote{\url{https://www.skatelescope.org/the-ska-project/}}),
\rev{the DSA-2000\footnote{\url{http://www.deepsynoptic.org}} \citep{2019BAAS...51g.255H}},
\rev{and the Canadian Hydrogen Observatory and Radio-transient Detector (CHORD) \citep{2019clrp.2020...28V}},
will further improve the detection rate of FRBs, making $\sim 10000$ 
events available in decades \citep[e.g.][]{2016MNRAS.460.1054C, 2020MNRAS.497.4107H}.

Because the DM contains all free electrons intervening between a given FRB and us, it is non-trivial how we can learn the redshift evolution of cosmic baryons from DM statistics, even if numerous FRBs become available.
To separate the observed DM into several pieces at different redshifts,
cross-correlation analyses with galaxies and galaxy clusters have been proposed \citep[e.g.][]{2014ApJ...780L..33M, 2017PhRvD..95h3012S, PhysRevD.100.103532, 2020PhRvD.102b3528R}.
There are two classes of cross-correlations: the angular correlation between the observed DM and spatial positions of galaxies (clusters)
\citep[e.g.][]{2014ApJ...780L..33M, 2017ApJ...834...13F, 2017PhRvD..95h3012S, PhysRevD.100.103532} and the correlation between spatial positions of FRB sources and galaxies (clusters)
\citep[e.g.][]{2017PhRvD..95h3012S, 2019ApJ...872...88R, 2020PhRvD.102b3528R}.
In this study, we work with the former cross-correlation because it is more sensitive to spatial distributions of electrons in and around gravitationally bound objects, referred to as dark matter haloes.
Notably, one can extract the information about cosmic free-electron distributions from the latter cross-correlation by dividing FRB sources into subsamples using their DM estimates \citep{2020PhRvD.102b3528R}.

The information contained in cross-correlation has not been explored in detail.
\citet{2014ApJ...780L..33M} showed that cross-correlation can be useful to statistically detect missing baryons around galaxies, whereas cosmological information in the cross-correlation has not been discussed.
\rev{\citet{2017ApJ...834...13F} considered DM-cluster correlation functions as a probe of missing baryons beyond cluster virial radii.}
\citet{2017PhRvD..95h3012S} proposed cross-correlations as a probe of redshift distributions and host environment of FRBs, but they ignored non-linear effects in the cross-correlation for simplicity.
\rev{\citet{2019ApJ...872...88R} studied a possibility of measuring the circumgalactic and intergalactic baryons with a sample of well-localized FRBs.}
\citet{PhysRevD.100.103532} focused on using FRBs as a probe of gas densities around galaxies, which is a key quantity in another cosmological observable of the kinematic Sunyaev-Zel'dovich (SZ) effect \citep{1980MNRAS.190..413S}.
\citet{2020PhRvD.102b3528R} have shown a realistic forecast of the detectability of FRB-galaxy cross-correlations in ongoing and future surveys, whereas it is still unclear what physics we can learn from a precise measurement of the cross-correlation. Actual analyses with available FRBs and galaxies have already shown a marginal detection of the cross-correlation \citep{2021ApJ...922...42R, 2021arXiv210713692C}.
Hence, it is timely to study the information content of cross-correlations as probes of missing baryons and cosmology.

In this study, we show that the precise measurement of cross-correlations 
constrains gastrophysics and cosmology separately (by breaking the degeneracy).
We improve the previous analyses of cross-correlation using a flexible and efficient physical model of cosmic gas density around a given dark matter halo \citep{2020MNRAS.495.4800A} and clarify possible degeneracy between cosmological parameters and assumed baryonic physics.
We adopt a halo-model approach \citep[e.g.][]{2002PhR...372....1C}
to predict cross-correlation and its statistical uncertainty for a given survey configuration.
Then, we study the information contents of the cross-correlation in terms of an expected signal-to-noise ratio (SNR) and a forecast of the expected constraints of relevant physical parameters via Fisher analysis.
Using the Fisher forecast, we find that cross-correlation allows 
us to place a tight constraint on cosmological parameters such as the mass variance at linear scales,
the average number density of free electrons in intergalactic media, 
and gas-to-halo mass relation from galaxy- to cluster-sized haloes.

The rest of this paper is organised as follows. 
We introduce the observables of interest and the theory of the cross-correlation analysis in Section~\ref{sec:basics}.
\en{Next}, we describe our model of gas density and the cross-correlation based on the halo model in Section~\ref{sec:model}.
The survey configuration in the Fisher forecast is summarised in Section~\ref{sec:setup}.
Section~\ref{sec:results} presents the key results, whereas we discuss the limitations of our analysis in Section~\ref{sec:limitations}.
Finally, concluding remarks are provided in Section~\ref{sec:conclusions}.
In the following, $\ln$ and $\log$ represent the natural logarithm and logarithm to base $10$, respectively.
We also provide a list of variables used in this paper in Table~\ref{tab:variables}.

\section{Preliminaries}\label{sec:basics}

\en{In} this \en{study}, we work with an expanding flat geometry universe described \en{using the} Friedmann-Robertson-Walker (FRW) metric.
Under general relativity, late-time expansion of the universe is governed by \en{the following equation:}
\beq
H(z)=H_{0}\sqrt{\Omega_\mathrm{M}(1+z)^3+\Omega_\mathrm{DE}(1+z)^{3(1+w_\mathrm{0})}}, \label{eq:Hz}
\eeq
where 
$z$ is the redshift, $H(z)$ is the Hubble parameter and
$H_0=100\, h\, \mathrm{km}/\mathrm{s}/\mathrm{Mpc}$ is the present value of $H(z)$.
In Eq.~(\ref{eq:Hz}), $\Omega_\mathrm{M}$ and $\Omega_\mathrm{DE}$ are 
the dimensionless density parameter of cosmic matter and dark energy (DE), respectively.
It holds that $\Omega_\mathrm{DE} = 1-\Omega_\mathrm{M}$ for the flat geometry.
We introduce an equation-of-state parameter of DE $w_0$ and simply assume no redshift dependence of the equation-of-state for DE.
\en{Notably}, $w_0=-1$ sets the DE to a cosmological constant $\Lambda$.
For a given $H(z)$, the radial comoving distance to redshift $z$ is computed as
\beq
\chi(z) = \int_0^{z}\!\frac{c\, \mathrm{d}z'}{H(z')}, \label{eq:chi_z}
\eeq
where $c$ is the speed of light. 
In the following, we will use $z$ and $\chi$ interchangeably.
 
\subsection{Projected number density field of haloes}

Let us assume that we have a sample of haloes distributed over a solid angle of the survey field. 
The angular number density of haloes per unit steradian can be written as \en{follows:}
\beq
n^{\rm 2D}_{\rm h}(\bmtheta)=\bar{n}^{\rm 2D}_{\rm h}\left[1+\delta_{\rm h}^{\rm 2D}\!(\bmtheta)\right], \label{eq:nhalo_2D}
\eeq
where $\delta_{\rm h}^{\rm 2D}(\bmtheta)$ is the projected number density fluctuation field, 
which is dimension-less. 
$\bar{n}^{\rm 2D}_{\rm h}$ is the mean number density, expressed in terms of the halo mass function as \en{follows:}
\beq
\bar{n}^{\rm 2D}_{\rm h}=\int_{\chi_{l,\mathrm{min}}}^{\chi_{l,\mathrm{max}}}\!\mathrm{d}\chi~ \chi^2 \int\!\!\mathrm{d}M~\frac{\mathrm{d}n}{\mathrm{d}M}(M,\chi) S(M,\chi).
\eeq
where
$\mathrm{d}n/\mathrm{d}M$ is the halo mass function, and $S(M,\chi)$ is the selection function of halo mass.
We introduce a specific form of $S$ in \en{Subsection}~\ref{subsec:foreground_haloes}.
We here set the range of comoving distances of interest to 
$\chi_{l,\mathrm{min}} \le \chi\le \chi_{l,\mathrm{max}}$.
The 2D field $\delta^{\rm 2D}_{\rm h}(\bmtheta)$ is expressed in terms of the 3D number density field of haloes as \en{follows:}
\beqa
\delta^{\rm 2D}_{\rm h}(\bmtheta)\equiv 
\frac{1}{\bar{n}^{\rm 2D}_{\rm h}}\int_{\chi_{l,\mathrm{min}}}^{\chi_{l,\mathrm{max}}}\!\!\mathrm{d}\chi~ \chi^2
\int\!\!\mathrm{d}M~\frac{\mathrm{d}n}{\mathrm{d}M}(M,\chi) \, S(M,\chi)   \nonumber  \\
 \times ~\delta_{\rm h}(\chi\bmtheta,\chi;M). \label{eq:delta_h_2D}
\eeqa
Here, we introduce the 3D number density fluctuation field of haloes, $\delta_{\rm h}$, via $n_{\rm h}(\bx; M) = ({\rm d}n/{\rm d}M)\, \left[1+\delta_{\rm h}(\bx; M)\right]$, where $n_{\rm h}(\bx; M)$ is the 3D number density field at the position, $\bx=(\chi\bmtheta, \chi)$, for haloes of mass $M$.

\subsection{Dispersion measure field}

In this \en{study}, we decompose the observed DM along a given line-of-sight into three components:
\beq
D_\mathrm{obs} = D_\mathrm{MW} + D_\mathrm{LSS} + D_\mathrm{source}, \label{eq:obs_DM}
\eeq
where $D_\mathrm{MW}$ represents contributions from the Milky Way, $D_\mathrm{source}$ \en{represents} contributions at the position of the FRB source of interest (basically from the host galaxy, 
see Subsection~\ref{subsec:D_source} for details), 
\en{and} $D_\mathrm{LSS}$ includes other contributions from intervening electrons between 
the source and observer.
\en{Notably}, LSS \en{means} large-scale structures, and $D_\mathrm{LSS}$ \en{comprises} circumgalactic, intergalactic and intracluster media.
In the following, we pay special attention to $D_\mathrm{LSS}$, which is expected to be a dominant contribution when the FRB source locates at 
extra-galactic distance scales.

Considering an FRB at an angular position $\bmtheta$ on the sky and redshift $z_s$, we formally express $D_\mathrm{LSS}$ as
\beq
D_\mathrm{LSS}(\bmtheta, z_s) = \int_0^{\chi_s}\!\mathrm{d}\chi~ n_\mathrm{e}(\chi\bmtheta, \chi)\,(1+z(\chi)),
\eeq
where $\chi_s = \chi(z_s)$, and $n_\mathrm{e}(\bx)$ is the comoving number density of free electrons.
The average number density of free electrons in intergalactic media is given by \citep{2014ApJ...783L..35D}
\beq
\bar{n}_\mathrm{e}(z) = \frac{\bar{\rho}_\mathrm{b}}{\mu_\mathrm{e} m_\mathrm{p}}\, f_\mathrm{e}(z), \label{eq:mean_ne}
\eeq
where 
$\bar{\rho}_\mathrm{b}$ is the cosmological baryon density in comoving units, $m_\mathrm{p}$ is the proton mass, $\mu_\mathrm{e} = (X_\mathrm{p} + Y_\mathrm{p}/2)^{-1}$ is the mean molecular weight of electrons,
$X_\mathrm{p}$ and $Y_\mathrm{p}$ represent the primordial mass fractions of hydrogen and helium, respectively.
\en{In} this \en{article}, we set $X_\mathrm{p} = 1-Y_\mathrm{p} = 0.76$.
In Eq.~(\ref{eq:mean_ne}), 
we introduce $f_\mathrm{e}(z)$, which describes the fraction of 
free electrons in the cosmic electron number density and
depends on redshift due to reionisation.
At a redshift $z_s$, $D_\mathrm{LSS}$ is \en{expressed as follows:}
\beqa
D_\mathrm{LSS}(\bmtheta, z_s) \!\!\!\!&=&\!\!\!\! \int_0^{\chi_s}\!\mathrm{d}\chi~ W_\mathrm{e}(\chi)\, 
\left[1+\delta_\mathrm{e}(\chi\bmtheta, \chi)\right], \label{eq:D_LSS_singlez}\\
W_\mathrm{e}(\chi) \!\!\!\!&\equiv&\!\!\!\! \frac{\bar{\rho}_\mathrm{b}}{\mu_\mathrm{e} m_\mathrm{p}}\, f_\mathrm{e}(z(\chi))\, \left(1+z(\chi)\right), \label{eq:W_e}
\eeqa
where $\delta_\mathrm{e}$ is the 3D number density fluctuation field of free electrons.
Eqs.~(\ref{eq:D_LSS_singlez}) and (\ref{eq:W_e}) show that $D_\mathrm{LSS}$ scales with $\Omega_\mathrm{b}h$ where $\Omega_\mathrm{b}$ is the dimensionless baryon density because $\bar{\rho}_\mathrm{b}$ varies with $\Omega_\mathrm{b}h^2$ and
the integration in Eq.~(\ref{eq:D_LSS_singlez}) gives an additional scaling of $h^{-1}$.

In reality, FRBs would follow a wide distribution in redshifts.
In this case, the DM-field $D_\mathrm{LSS}$ is \en{expressed as follows} \citep{2017PhRvD..95h3012S}:
\beqa
D_\mathrm{LSS}(\bmtheta) \!\!\!\!&=&\!\!\!\! \int_0^{\chi_H}\, \!\mathrm{d}\chi~ K_\mathrm{e}(\chi)\left[1+\delta_\mathrm{e}(\chi\bmtheta, \chi)\right], \label{eq:DM_IGM_widez} \\
K_\mathrm{e}(\chi) \!\!\!\!&\equiv&\!\!\!\! W_\mathrm{e}(\chi)\, \int_{\chi}^{\chi_H}\!\mathrm{d}\chi'~ p(\chi'),
\eeqa
where $\chi_H$ is the comoving distance to $z\rightarrow \infty$
and $p(\chi)$ represents the redshift distribution of FRBs.
\en{Notably}, we reverse the order of integration in $\chi$ and $\chi'$ to derive Eq.~(\ref{eq:DM_IGM_widez}).
We adopt the following functional form of $p(\chi)$:
\beqa
p(\chi) \!\!\!\!&=&\!\!\!\! p_z(z)\, \left(\frac{\mathrm{d}\chi}{\mathrm{d}z}\right)^{-1}, \\
p_z(z) \!\!\!\!&=&\!\!\!\! p_0\, z^2\, \exp(-\alpha z)\, \Theta(z_\mathrm{max, f}-z)\, \Theta(z), \label{eq:pz}
\eeqa
where $\Theta(x)$ is the Heaviside step function, $\alpha$ and $z_\mathrm{max,f}$ are free parameters in the model of $p(\chi)$.
We here assume a simple form of Eq.~(\ref{eq:pz}) 
to mimic an actual redshift distribution of FRBs, but our cross-correlation analysis does not require a precise determination of individual FRB redshifts.
Our cross-correlation analysis uses the redshift information from
other galaxy observations, allowing us to constrain the FRB redshift distribution.
In this paper, our baseline model assumes $\alpha=3.5$ and $z_\mathrm{max,f} = 5.0$, which set the median redshift to $0.577$.
The current observation suggests that 
a major population of FRBs should have
the source redshift less than unity \citep{2021arXiv210604352T}.
Notably, we set the normalisation $p_0$ in Eq.~(\ref{eq:pz}) by
imposing $\int_0^{\infty}\mathrm{d}z\, p_z(z)=1$.
Our baseline model of the redshift distribution largely corresponds to 
the scenario of FRBs following the star-formation history \citep[e.g.][]{2016PhRvL.117i1301M}, whereas the realistic values of $\alpha$ and $z_\mathrm{max,f}$ depends on the properties of survey instruments and the origin of FRBs.
When making a forecast of parameter constraints, we vary the parameter of $\alpha$ to marginalise the uncertainty of the redshift estimation.

Although the redshift distribution of FRBs is uncertain at present, 
we find that our Fisher analysis in Subsection~\ref{subsec:Fisher_results} is less sensitive to the choice of $\alpha$ as long as we vary the fiducial $\alpha$ by $\pm1$.
Note that $\alpha=2.5$ and $\alpha=4.5$ set the median redshifts of $0.80$ and $0.44$, respectively.
The expected cosmological constraints can be affected by the change of $\alpha=3.5\pm1$ with a level of $\sim1\%$.

\subsection{Cross-correlations}

Our primary focus is to extract the information about cosmic free electrons from cross-correlation analyses between the projected halo density field $\delta^\mathrm{2D}_\mathrm{h}$ and the \en{DM-field} $D_\mathrm{obs}$. 
First, we define a two-point cross-correlation function as \en{follows:}
\beq
\xi_{\mathrm{h}\mathrm{D}}(\theta) 
\equiv
\langle \delta^\mathrm{2D}_\mathrm{h}(\bmtheta_1) D_\mathrm{obs}(\bmtheta_2)\rangle
\simeq
\langle \delta^\mathrm{2D}_\mathrm{h}(\bmtheta_1) D_\mathrm{LSS}(\bmtheta_2)\rangle \label{eq:cross_2pcf}
\eeq
where $\langle \cdots \rangle$ represents an ensemble average, 
$\theta \equiv |\bmtheta_1-\bmtheta_2|$, and 
we ignore possible correlations among $\delta^\mathrm{2D}_\mathrm{h}$, $D_\mathrm{MW}$ 
and $D_\mathrm{source}$.
It holds that $\langle\delta^\mathrm{2D}_\mathrm{h}\, D_\mathrm{MW}\rangle=0$ because we consider the halo sample at extra-galactic scales.
On the other hand, there may exist correlations between $\delta^\mathrm{2D}_\mathrm{h}$ and $D_\mathrm{source}$ \en{provided} some FRBs occur inside the haloes of interest \citep[e.g.][]{2017PhRvD..95h3012S}.
Nevertheless, the correlation of $\langle\delta^\mathrm{2D}_\mathrm{h}\, D_\mathrm{source}\rangle$ is largely uncertain at present because it is related to the origin of FRBs.
Notably, $D_\mathrm{source}$ has been estimated to be 
$\simlt 200\, \mathrm{pc}/\mathrm{cm}^3$ for FRB host galaxies with known redshifts, except for FRB 190520 \citep[e.g.][]{2016Natur.531..202S, 2017Natur.541...58C, 2017ApJ...834L...7T, 2017ApJ...844...95K, 2017ApJ...843L...8B, 2019Sci...366..231P, 2019Natur.572..352R, 2019Sci...365..565B, 2020Natur.582..351C, 2020arXiv200513158C, 2020ApJ...895L..37B, 2020ApJ...901L..20B, 2020arXiv201211617M, 2020Natur.577..190M, 2020Natur.581..391M, 2020ApJ...901..134S, 2020ApJ...899..161L, 2020ApJ...903..152H, 2021arXiv210107998J, 2021ApJ...910L..18B}.
Hence, we simply ignore the term of $\langle\delta^\mathrm{2D}_\mathrm{h}\, D_\mathrm{source}\rangle$ 
in the cross-correlation function. 
We briefly discuss the impact of $\langle\delta^\mathrm{2D}_\mathrm{h}\, D_\mathrm{source}\rangle$ on our analysis in Subsection~\ref{subsec:D_source}, whereas we leave the investigation of $\langle\delta^\mathrm{2D}_\mathrm{h}\, D_\mathrm{source}\rangle$ for future studies.

The cross-correlation in Fourier space is called the cross power spectrum\footnote{In this paper, we work with a flat-sky approximation. The power spectrum is defined in Fourier space throughout our paper. Hence, a multipole $\bmell$ consists of real numbers. In Eq.~(\ref{eq:xi2Cl}), the norm of multipole $\ell=|\bmell|$ 
is loosely related to an angular separation $\theta$ by
$\ell = 1/\theta \simeq 3437\, (\theta/1\, \mathrm{arcmin})^{-1}$.}, 
which is defined as follows:
\beq
\langle \tilde{\delta}^\mathrm{2D}_\mathrm{h}(\bmell_1)\tilde{D}_\mathrm{LSS}(\bmell_2) \rangle 
\equiv (2\pi)^2\, \delta^{(2)}_\mathrm{D}(\bmell_1+\bmell_2)\, C_{\mathrm{hD}}(\ell_1), \label{eq:def_Cl}
\eeq
where 
$C_\mathrm{hD}$ is the cross power spectrum, $\delta^{(n)}_\mathrm{D}(\bx)$ is the $n$-dimensional Dirac delta function, $\tilde{\delta}^\mathrm{2D}_\mathrm{h}$ is the Fourier transform of $\delta^\mathrm{2D}_\mathrm{h}$ and so on. 
The cross power spectrum is equivalent to the Fourier transform of $\xi_\mathrm{hD}$, and it holds that
\beq
\xi_\mathrm{hD}(\theta) = \int \frac{\mathrm{d}^2\ell}{(2\pi)^2}\, C_\mathrm{hD}(\ell) \,
\exp(i\bmell\cdot\bmtheta). \label{eq:xi2Cl}
\eeq

Then, we relate the two-point correlation function $\xi_\mathrm{hD}$ with clustering information of underlying 3D fields $\delta_\mathrm{h}(\bx,M)$ and $\delta_\mathrm{e}(\bx)$.
Using Eqs.~(\ref{eq:delta_h_2D}) and (\ref{eq:DM_IGM_widez}), \en{we obtain}
\beqa
\xi_\mathrm{hD}(\theta) \!\!\!\!&=&\!\!\!\! 
\frac{1}{\bar{n}^\mathrm{2D}_\mathrm{h}} \int_{\chi_{l,\mathrm{min}}}^{\chi_{l,\mathrm{max}}}\!\mathrm{d}\chi_1\, \chi^2_1\,
\int_0^{\chi_H}\!\mathrm{d}\chi_2\, K_\mathrm{e} (\chi_2) \nonumber \\
&\times&\!\!\!\!
\int\! \mathrm{d}M\, \frac{\mathrm{d}n}{\mathrm{d}M}(M,\chi_1)  \, S(M, \chi_1)
\int\!\frac{\mathrm{d}^3k}{(2\pi)^3}\, P_{\mathrm{he}}(k, M, \chi_1, \chi_2) \nonumber \\
&\times&\!\!\!\!
\exp\left[i\bk_\perp \cdot \left(\chi_1\bmtheta_1 - \chi_2 \bmtheta_2\right)\right]
\exp\left[ik_\parallel \left(\chi_1-\chi_2\right)\right], \label{eq:xi_hD_full}
\eeqa
where $\bk = (\bk_\perp, k_\parallel)$, we perform the Fourier transform of the 3D fields of $\delta_\mathrm{h}$ and $\delta_\mathrm{e}$, and define the 3D power spectrum as
\beq
\langle \tilde{\delta}_\mathrm{h}(\bk_1, M,\chi_1) \tilde{\delta}_\mathrm{e}(\bk_2, \chi_2)\rangle 
\equiv (2\pi)^3\, \delta^{(3)}_\mathrm{D}(\bk_1+\bk_2)\, P_{\mathrm{he}}(k_1, M, \chi_1, \chi_2).
\label{eq:P_he}
\eeq
We adopt the Limber approximation \citep{1954ApJ...119..655L} so that
\beqa
&&\!\!\!\!\int\frac{\mathrm{d}k_\parallel}{2\pi}\, P_{\mathrm{he}}(k, M, \chi_1, \chi_2)\, \exp\left[ik_\parallel \left(\chi_1-\chi_2\right)\right] \nonumber \\
&\simeq&\!\!\!\!
P_{\mathrm{he}}(k_\perp, M, \chi_1, \chi_2) \, \delta^{(1)}_D(\chi_1-\chi_2).
\eeqa
Using Eq.~(\ref{eq:xi2Cl}), we \en{obtain}
\beqa
C_\mathrm{hD}(\ell) \!\!\!\!&=&\!\!\!\!
\frac{1}{\bar{n}^\mathrm{2D}_\mathrm{h}}\int_{\chi_{l,\mathrm{min}}}^{\chi_{l,\mathrm{max}}} \!\mathrm{d}\chi\, K_\mathrm{e}(\chi)
\int \!\mathrm{d}M\, \frac{\mathrm{d}n}{\mathrm{d}M}(M,\chi)\, S(M,\chi) \nonumber \\
&&\!\!\!\!
\times
P_{\mathrm{he}}\left(k = \frac{\ell}{\chi}, M, \chi, \chi \right), \label{eq:Cl_limber}
\eeqa
where we use $\bk_\perp \chi = \bmell$ to derive Eq.~(\ref{eq:Cl_limber}).
In the following, we study the cross power spectrum $C_\mathrm{hD}$
to extract the clustering information of cosmic free electrons around a given sample of haloes
and cosmological parameters.
For a simpler notation, we write $P_{\mathrm{he}}\left(k = \ell/\chi, M, \chi, \chi \right)$
as $P_{\mathrm{he}}\left(k = \ell/\chi, M, \chi\right)$ below.

\section{Model of free electrons}\label{sec:model}

\subsection{Halo model}

To compute the cross power spectrum $C_\mathrm{hD}$, we need the model of the 3D power spectrum of $P_{\mathrm{he}}$ defined in Eq.~(\ref{eq:P_he}).
We \en{adopt} a halo-model approach \citep{2002PhR...372....1C, 2020PhRvD.102b3528R} to compute $P_{\mathrm{he}}$.

The halo model assumes that all free electrons are associated with single haloes.
Under this assumption, the power spectrum can be decomposed into two parts:
\beq
P_\mathrm{he}(k, M, z) =  P^{1\mathrm{h}}_\mathrm{he}(k, M, z) + P^{2\mathrm{h}}_\mathrm{he}(k, M, z), \label{eq:halo_model}
\eeq
where $z$ is the redshift corresponding to a given comoving distance $\chi$.
The former and latter \en{on} the right-hand side in Eq.~(\ref{eq:halo_model}) are 1- and 2-halo terms, respectively.
The 1-halo term arises from the two-point correlation in single haloes, whereas the 2-halo term expresses the two-point correlation function between neighbouring haloes.
Considering that free electrons around a halo of $M$ follow a spherical density profile of $n_\mathrm{e,h}(r, M, z)$, \en{we obtain}
\beqa
P^{1\mathrm{h}}_\mathrm{he}(k, M, z) \!\!\!\!&=&\!\!\!\!
\frac{\tilde{n}_\mathrm{e,h}(k, M, z)}{\bar{n}_\mathrm{e,hm}(z)}, \label{eq:P1h}\\
P^{2\mathrm{h}}_\mathrm{he}(k, M, z) \!\!\!\!&=&\!\!\!\!
\int \!\mathrm{d}M'\, \frac{\mathrm{d}n}{\mathrm{d}M'}(M',z)
\frac{\tilde{n}_\mathrm{e,h}(k, M', z)}{\bar{n}_\mathrm{e,hm}(z)} \nonumber \\
\!\!\!\!&&\!\!\!\!
\quad \quad \quad
\quad \quad \quad
\times \, P_\mathrm{hh}(k, M, M', z), \label{eq:P2h}\\
\bar{n}_\mathrm{e,hm}(z) \!\!\!\!&=&\!\!\!\! \int \!\mathrm{d}M\, \frac{\mathrm{d}n}{\mathrm{d}M}(M,z)\, 
\int \!\mathrm{d}V\, n_\mathrm{e,h}(r, M, z), \label{eq:hm_norm}
\eeqa
where $\tilde{n}_\mathrm{e,h}$ is the Fourier transform of the density profile $n_\mathrm{e,h}(r, M, z)$, and $P_\mathrm{hh}(k,M,M',z)$ is 
the 3D power spectrum between $\delta_\mathrm{h}(\bx,M)$
and $\delta_\mathrm{h}(\bx,M')$ at redshift $z$.
We adopt the linear approximation for $P_\mathrm{hh}(k,M,M',z)$:
\beq
P_\mathrm{hh}(k, M, M',z) \simeq b_\mathrm{h}(M, z)b_\mathrm{h}(M', z)P_\mathrm{L}(k,z), \label{eq:2halo_hh}
\eeq
where $b_\mathrm{h}$ is the linear halo bias, and $P_\mathrm{L}$ represents the linear matter power spectrum.

To specify our halo model, we set the halo mass of $M$ by a spherical overdensity mass at $200$ times the critical density, referred to as $M_\mathrm{200c}$ in the literature.
For this mass, we adopt the model of the halo mass function in \citet{2008ApJ...688..709T} and the linear halo bias in \citet{2010ApJ...724..878T}.
\en{In addition}, we compute the linear matter power spectrum using a Boltzmann code \citep[{\tt CAMB}, see][for details]{Lewis:1999bs}.
The key ingredient \en{of} our model, $n_\mathrm{e,h}$, is set in the next subsection.

\subsection{Two-component gas model}\label{subsec:gas_model}

\begin{table*}
 \caption{A short summary of the model parameters of stars and gas in \citet{2020MNRAS.495.4800A}.
 The parameter $\eta$ is a typical distance scale of a gas particle expelled by the feedback 
 (in units of $5.30 \, r_\mathrm{200c}$),
 $M_c$ gives the halo mass for which 50\% of the gas in haloes is in the hot and bound state,
 $\beta$ controls how rapidly the gas mass in the hot state decreases as reducing the halo mass,
 and $M_{1,0}$ is the pivot halo mass giving the mass fraction of stars being 0.023 at $z=0$.
 We list the inferred value of each parameter in \citet{2020MNRAS.495.4800A} for different hydrodynamical simulations:
 TNG300 \citep{2018MNRAS.475..676S}, BAHAMAS \citep{2017MNRAS.465.2936M}, 
 Cosmo-OWLS \citep{2014MNRAS.441.1270L}, Horizon-AGN \citep{2014MNRAS.444.1453D}
 and EAGLE \citep{2015MNRAS.446..521S, 2016MNRAS.461L..11H}.
 Note that we adopt the best-fit parameters of $\eta$, $M_c$ and $\beta$ for the simulations at $z=0$ and ignore a mild redshift dependence throughout the paper.
 }
 \label{tab:model_summary}
 \begin{tabular}{lccccc}
  \hline
  Parameters & TNG300 & BAHAMAS & Cosmo-OWLS & Horizon-AGN & EAGLE \\
  \hline
  $\eta$                              & 0.14 & 0.53 & 0.35 & 0.15 & 0.14 \\
  $M_c\, [10^{13} h^{-1}M_\odot]$     & 2.3 & 3.8 & 0.4 & 1.2 & 1.8 \\
  $\beta$                             & 4.09 & 0.47 & 0.25 & 6.38 & 9.65 \\
  $M_{1,0}\, [10^{11} h^{-1}M_\odot]$ & 0.22 & 10.85 & 1.61 & 0.07 & 11.15 \\
  \hline
 \end{tabular}
\end{table*}

For the number density profile of electrons around a halo, 
we adopt the model developed in \citet{2015JCAP...12..049S, 2020MNRAS.495.4800A}.
The model assumes that gas in a given halo of $M$ \en{comprises} two components: a hot gas, assumed to be in hydrostatic equilibrium inside haloes, and a gas ejected from the halo due to some feedback processes 
by, e.g. active galactic nuclei (AGN).

We consider four free parameters for the gas model: a typical distance scale of ejected gas from the halo centre ($\eta$), a characteristic halo mass for which 50\% of the gas in a halo is in the hot bound state ($M_c$), mass dependence of the gas depletion by the feedback ($\beta$), and a characteristic halo mass providing the stellar mass fraction of 0.023 at $z=0$ ($M_{1,0}$).
\citet{2020MNRAS.495.4800A} calibrated \en{these} parameters with a set of hydrodynamical simulations so that the model can reproduce the power spectrum of the matter density fields in the simulations.
\en{Notably}, direct calibrations of gas density profiles \en{were not performed} in \citet{2020MNRAS.495.4800A}.

Hence, the model is still phenomenological and not accurate enough for some purposes 
(e.g. the application to real data in the future).
Nevertheless, we find that the halo-model prediction with the gas model in \citet{2020MNRAS.495.4800A} can explain cross power spectra based on the TNG simulation \citep{2018MNRAS.475..676S} within a level of $\sim 25\%$--$30\%$ at a wide range of halo masses, redshifts and multipoles $\ell$ (also see Appendix~\ref{apdx:comp_TNG}).
We believe that our model is sufficient to study information contents in the cross power spectra of $C_\mathrm{hD}$ at this early stage.
We leave more careful modelling of $C_\mathrm{hD}$ for future studies.

Table~\ref{tab:model_summary} \en{lists} our model parameters and the inferred values from the analysis in \citet{2020MNRAS.495.4800A} for different hydrodynamical simulations.

Apart from the parameters for gas density, we also introduce two nuisance parameters
to marginalise over uncertainties in baryonic effects in total matter density profiles in single haloes ($A_\mathrm{DK15}$ in Eq.~[\ref{eq:c_h}])
and the amplitude in the fraction of free electrons in cosmic baryon density
($A_\mathrm{e,norm}$ in Eq.~[\ref{eq:f_e_model}]).

\subsubsection{Bound gas}

For the hot gas component, the mass density profile is modelled as
\beq
\rho_\mathrm{BG}(r) = f_\mathrm{BG}\, M
\left\{
\begin{array}{ll}
y_0\, \left[x^{-1}\ln(1+x)\right]^{\Gamma} & r<r_\mathrm{200c}/\sqrt{5} \\
y_1\, x^{-1}(1+x)^{-2} & r_\mathrm{200c}/\sqrt{5} \le r < r_\mathrm{200c} \\
0 & r\ge r_\mathrm{200c}
\end{array}
\right., \label{eq:rho_BG}
\eeq
where $r_\mathrm{200c}$ is given by $r_\mathrm{200c} = (1+z)\, [3M/(4\pi\,\rho_\mathrm{crit}(z))/200]^{1/3}$ and $\rho_\mathrm{crit}(z)=3\, H^2(z)/(8\pi\, G)$ is the critical density in the universe at redshift $z$, $x=r/r_s$ and $r_s$ is the scale radius of a spherical Navarro-Frenk-White (NFW) profile for dark matter haloes \citep{1996ApJ...462..563N}.
\en{Notably}, we work with the comoving coordinate.
To compute the scale radius, we introduce a halo concentration parameter as $c_\mathrm{h} = r_\mathrm{200c}/r_s$. 
We adopt the model of $c_\mathrm{h}$ in \citet{2015ApJ...799..108D} 
with a free normalisation, i.e.
\beq
c_\mathrm{h}(M,z) = A_\mathrm{DK15}\, c_\mathrm{DK15}(M,z), \label{eq:c_h}
\eeq
where $A_\mathrm{DK15}$ is a free parameter and $c_\mathrm{DK15}$ is the model in \citet{2015ApJ...799..108D}.
\en{Notably}, $c_\mathrm{DK15}$ depends on the amplitude and shape of the linear matter power spectrum $P_\mathrm{L}$ \citep[see][for details]{2015ApJ...799..108D}.
Hence, the concentration shows a cosmological dependence; it is also affected by baryonic feedback processes in single haloes \citep[e.g.][]{2010MNRAS.405.2161D}.
Thus, we include the free parameter $A_\mathrm{DK15}$ to marginalise the baryonic effects in $c_\mathrm{h}$.
The inner slope of $\Gamma$ in Eq.~(\ref{eq:rho_BG}) is defined such that the hydrostatic gas has the same slope of the NFW at $r=r_\mathrm{200c}/\sqrt{5}$.
To be specific, it is given by
\beq
\Gamma(M,z) = \frac{\left(1+3c_\mathrm{h}/\sqrt{5}\right)\ln\left(1+c_\mathrm{h}/\sqrt{5}\right)}{\left(1+c_\mathrm{h}/\sqrt{5}\right)\ln\left(1+c_\mathrm{h}/\sqrt{5}\right)-c_\mathrm{h}/\sqrt{5}}.
\eeq
The normalisation factors $y_0$ and $y_1$ are set so that the profile can be continuous and $\int 4\pi r^2 \mathrm{d}r \rho_\mathrm{BG}(r) = f_\mathrm{BG}\, M$.
We assume that the gas fraction $f_\mathrm{BG}$ depends on the mass $M$ and redshifts $z$:
\beq
f_\mathrm{BG}(M,z) = \frac{\Omega_\mathrm{b}/\Omega_\mathrm{M}-f_\mathrm{star}(M,z)}{1+\left(M_c/M\right)^{\beta}}, \label{eq:frac_BG}
\eeq
where 
$f_\mathrm{star}$ describes the stellar-to-halo mass ratio,
$M_c$ and $\beta$ are free parameters in the model.
The explicit form of $f_\mathrm{star}$ is given in Appendix~\ref{apdx:f_star}; it contains a single free parameter $M_{1,0}$ giving $f_\mathrm{star}=0.023$ at $M=M_{1,0}$ and $z=0$.

\subsubsection{Ejected gas}

The gas ejected from the halo is assumed to follow the density distribution below:
\beq
\rho_\mathrm{EG}(r) = \frac{f_\mathrm{EG}M}{(2\pi r^{2}_\mathrm{ej})^{3/2}}
\exp\left[-\frac{1}{2}\left(\frac{r}{r_\mathrm{ej}}\right)^2\right], \label{eq:rho_EG}
\eeq
where the profile is derived \en{via} a Maxwell-Boltzmann velocity distribution of the particles expelled by the AGN under simple circumstances \citep{2015JCAP...12..049S}.
The radius $r_\mathrm{ej}$ is a parameter in the model.
For convenience, we introduce a dimensionless parameter for $r_\mathrm{ej}$:
\beq
r_\mathrm{ej} = 0.75\, \eta\, r_\mathrm{esc}, 
\eeq
where $\eta$ is the model parameter and $r_\mathrm{esc}$ is the halo escape radius.
We estimate $r_\mathrm{esc}$ by the distance travelled by a gas particle with a constant halo escape velocity over a time-scale of a half Hubble time. 
For the mass of $M$, $r_\mathrm{esc}$ is given by $7.07\, r_\mathrm{200c}$.
The mass fraction of ejected gas is simply set \en{as follows:}
\beq
f_\mathrm{EG}(M,z) = \Omega_\mathrm{b}/\Omega_\mathrm{M} - f_\mathrm{BG}(M,z) - f_\mathrm{star}(M,z). \label{eq:frac_EG}
\eeq

\subsubsection{Conversion of gas into \rev{electrons}}

For given density profiles of $\rho_\mathrm{BG}$ and $\rho_\mathrm{EG}$,
we convert the gas mass density into the number density of electrons with a constant factor:
\beq
n_\mathrm{e,h}(r,M,z) = \frac{\rho_\mathrm{BG}(r,M,z) + \rho_\mathrm{EG}(r,M,z)}{\mu'_\mathrm{e}(z)m_\mathrm{p}},
\eeq
where 
$\mu'_\mathrm{e}(z)$ is an effective molecular weight of electrons at different $z$.
We simply set $\mu'_\mathrm{e}(z)=(X_\mathrm{p}+Y_\mathrm{p}/2)^{-1}$.
\en{Notably}, our results are insensitive to the choice of $\mu'_\mathrm{e}(z)$ 
because the normalisation is cancelled in the halo model (see Eqs.~[\ref{eq:P1h}]-[\ref{eq:hm_norm}]).

\subsection{Fraction of free electrons in cosmic baryons}

The overall amplitude of $C_\mathrm{hD}$ depends on the fraction of free electrons in baryon density, $f_\mathrm{e}$ in Eq.~(\ref{eq:mean_ne}).
At high redshifts ($z\simgt 6$), the gas is expected to still be neutral.
At lower redshifts, the fraction of free electrons would rapidly grow because of hydrogen reionisation and increases further at the epoch of helium reionisation. 
Even after the helium reionisation, the fraction can decrease 
because some fraction of the electrons may be confined in stars and black holes.

We use a realistic model of $f_\mathrm{e}(z)$, which has been calibrated in \citet{2021MNRAS.502.2615T}. The model is given by
\beq
f_\mathrm{e, T21}(z) = 0.475 (z+0.703)^{0.02}\left[1-\tanh\{3.19\left(z-5.42\right)\}\right], \label{eq:f_e_fid}
\eeq
where it can reproduce the fraction of free electrons in the TNG simulation.
To make our halo model self-consistent, we include \en{the} possible dependence of $f_\mathrm{e}$ on gas-to-star conversion and possible uncertainties in Eq.~(\ref{eq:f_e_fid}). 
Our final model of $f_\mathrm{e}(z)$ is then given by
\beq
f_\mathrm{e}(z) = 
A_\mathrm{e,norm}\, f_\mathrm{e, T21}(z)\, \frac{\rho_\mathrm{tot, gas}(z; M_{1,0})}{\rho_\mathrm{tot, gas}(z; M_{1,0})|_\mathrm{TNG}}, \label{eq:f_e_model}
\eeq
where 
$A_\mathrm{e,norm}$ is a free parameter in the model,
and we define
\beq
\rho_\mathrm{tot, gas}(z; M_{1,0}) = 
\int \!\mathrm{d}M\, \frac{\mathrm{d}n}{\mathrm{d}M}(M,z) \, M \,
\left[\frac{\Omega_\mathrm{b}}{\Omega_\mathrm{M}}-f_\mathrm{star}(M,z)\right],
\eeq
and $\rho_\mathrm{tot, gas}|_\mathrm{TNG}$ represents the result 
when we set $M_{1,0} = 2.2 \times 10^{10}\, h^{-1}M_\odot$.

\section{Analysis setup}\label{sec:setup}

In this section, we summarise the setup for cross-correlation 
analyses in an FRB hypothetical survey. We assume that halo/FRB catalogues covering a sky of $20000\, \mathrm{deg}^2$ are available.
\rev{This sky coverage can be available in upcoming/ongoing full-sky surveys of SPHEREx\footnote{\url{https://spherex.caltech.edu/}} \citep{2018arXiv180505489D}
and eROSITA\footnote{\url{https://www.mpe.mpg.de/eROSITA}} \citep[e.g.][]{2012arXiv1209.3114M}.}

\subsection{Foreground haloes}\label{subsec:foreground_haloes}

\begin{figure}
 \includegraphics[width=\columnwidth]{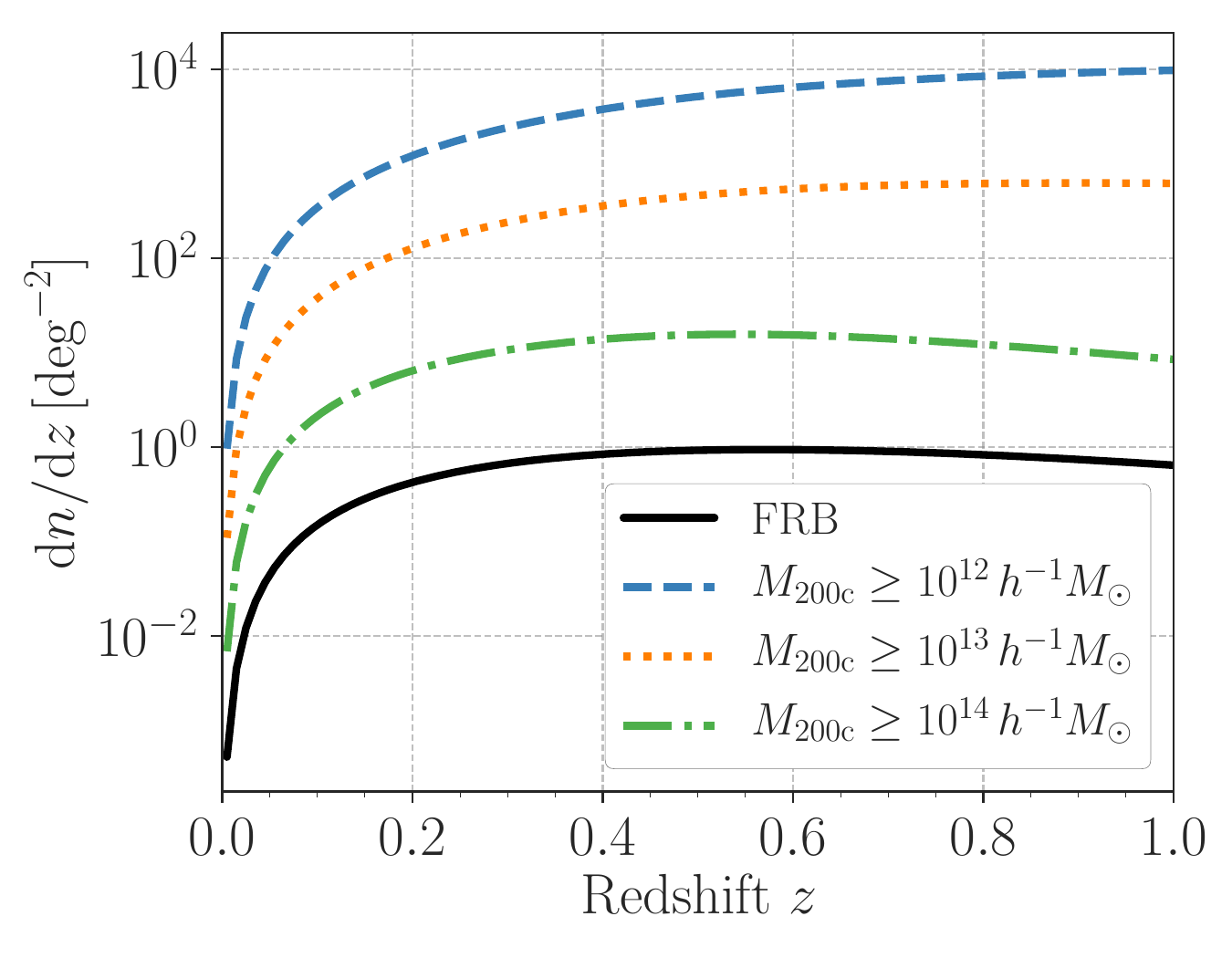}
 \caption{The redshift distribution of FRBs and foreground haloes in our fiducial analysis setup.
 }
 \label{fig:dndz}
\end{figure}

We consider an ideal setup of foreground halo samples to give some sense of how much information contents can be extracted from cross-correlation analyses.
For cross-correlations, we study three mass-limited samples with $M \ge M_\mathrm{thr}$ at five redshift bins.
As a representative example, we use three mass thresholds of 
$M_\mathrm{thr}\, [h^{-1}M_\odot] = 10^{12}$, $10^{13}$ and $10^{14}$.
The halo masses of $10^{12}$, $10^{13}$ and $10^{14}\, h^{-1}M_\odot$
largely correspond to the galaxy main sequence, red luminous galaxies and galaxy clusters, respectively \citep{2018ARA&A..56..435W}.
We further assume that the halo catalogue contains no satellite galaxies for simplicity.
In this setup, the mass selection function $S(M)$ is simply given by $S(M)=\Theta(M-M_\mathrm{thr})$.
For the redshift bins, we set the edge of each bin to 
$0<z\le0.2$, $0.2<z\le0.4$, $0.4<z\le 0.6$, $0.6<z\le 0.8$ and $0.8<z\le 1.0$.
We assume that the redshifts of individual haloes are precisely measured and do not \en{consider} any errors in their redshift measurements.
A more realistic model of galaxies and clusters based on, e.g. 
halo occupation distribution \citep[e.g.][]{2003ApJ...593....1B}
would be relevant to actual analyses, \en{but} we leave it for future studies.
Figure~\ref{fig:dndz} shows the redshift distribution of the foreground halo samples and FRBs.

\rev{It would be worth noting that halo catalogues with precise redshift measurements are likely available in the range of $0\simlt z\simlt 1$ in the future. 
On galaxy-sized haloes, 
there already exist wide-area catalogues of spectroscopic redshifts covering $z \simlt 0.7$, provided by the 6dF Galaxy Survey\footnote{\url{http://www.6dfgs.net/}} \citep[e.g.][]{2009MNRAS.399..683J}, 
the Sloan Digital Sky Survey (SDSS\footnote{\url{https://www.sdss.org/}}) \citep[e.g.][]{2007ApJS..172..634A, 2009ApJS..182..543A}, 
the SDSS's Baryon Oscillation Spectroscopic Survey (BOSS) \citep[e.g.][]{2013AJ....145...10D}
and the WiggleZ Dark Energy Survey\footnote{\url{https://wigglez.swin.edu.au/site/forward.html}} \citep[e.g.][]{2018MNRAS.474.4151D}. 
In addition, upcoming surveys can make catalogues of
spectroscopic redshifts at $0.7 \simlt z \simlt 1$ available. 
Those include the Subaru Prime Focus Spectrograph (PFS\footnote{\url{https://pfs.ipmu.jp/}}) \citep{2014PASJ...66R...1T}, 
the Dark Energy Spectroscopic Instrument (DESI\footnote{\url{https://www.desi.lbl.gov/}}) \citep{2016arXiv161100036D}, 
and SPHEREx.}

\rev{On cluster-sized haloes, the SDSS imaging data enable us to construct 
a wide-area catalogue of $\sim 10^{14}\, M_\odot$ haloes 
in the range of $0.1\simlt z \simlt 0.6$ with an empirical optical cluster finder \citep[e.g.][]{2014ApJ...785..104R, 2018PASJ...70S..20O}. 
Ongoing multi-wavelength surveys, e.g. the South Pole Telescope (SPT\footnote{\url{https://pole.uchicago.edu/public/Home.html}}) \citep[e.g][]{2020ApJS..247...25B}, 
the Atacama Cosmology Telescope (ACT\footnote{\url{https://act.princeton.edu/}}) \citep[e.g.][]{2018ApJS..235...20H} and eROSITA, can complete clusters with their mass of $\simgt 10^{14}\, M_\odot$ at $z \simlt 1$ with more secure finders. The redshifts of galaxy clusters can be calibrated with a sample of spectroscopic galaxies through cross matching on an object-by-object basis and/or cross-correlation analyses.}

\subsection{Properties of FRBs}

\en{For} the cross-correlation analysis, we have to set the average angular number density of FRBs $\bar{n}_\mathrm{FRB}$, the variance of DM associated with FRB host galaxies, denoted as $\sigma_\mathrm{DM}$,
and the angular resolution of FRB position's $\Delta \theta_\mathrm{FRB}$.
We assume that $D_\mathrm{MW}$ can be precisely subtracted from the observed DM and ignore possible small variance from the residual of $D_\mathrm{MW}$.
We also assume that the redshift distribution FRBs is given by Eq.~(\ref{eq:pz}).
For a given $\Delta \theta_\mathrm{FRB}$, we compute a smearing effect on the cross power spectrum due to the uncertainty of the FRB locations as \en{follows} \citep{2020PhRvD.102b3528R}:
\beqa
C_\mathrm{hD}(\ell) \!\!\!\!&\rightarrow&\!\!\!\! C_\mathrm{hD}(\ell)\, {\cal B}(\ell), \\
{\cal B}(\ell) \!\!\!\!&=&\!\!\!\! \exp\left(-\frac{(\Delta \theta_\mathrm{FRB}\ell)^2}{16\ln 2}\right), \label{eq:beam}
\eeqa
where $\Delta \theta_\mathrm{FRB}$ provides a full width at half maximum in the Gaussian probability distribution of angular positions for individual FRBs.

For our fiducial case, we set $\bar{n}_\mathrm{FRB}=1\, \mathrm{deg}^{-2}$, $\sigma_\mathrm{DM} = 60\, \mathrm{pc}/\mathrm{cm}^3$,
and $\Delta \theta_\mathrm{FRB}=3\, \mathrm{arcmin}$.
We set the value of $\sigma_\mathrm{DM} = 60\, \mathrm{pc}/\mathrm{cm}^3$ \en{following} the estimate of the probability distribution of $D_\mathrm{source}$ in \citet{2020Natur.581..391M}.
Our fiducial values of $\bar{n}_\mathrm{FRB}=1\, \mathrm{deg}^{-2}$ and $\Delta \theta_\mathrm{FRB}=3\, \mathrm{arcmin}$ can be realised in future \en{FRB} surveys \citep[e.g.][]{2018ApJ...863...48C}, \en{whereas} we will examine two other cases of $(\bar{n}_\mathrm{FRB}, \Delta \theta_\mathrm{FRB}) = (0.1\, \mathrm{deg}^{-2}, 1\, \mathrm{arcmin})$
and $(10\, \mathrm{deg}^{-2}, 10\, \mathrm{arcmin})$ as \en{required}.
Note that we test how the parameters of $\bar{n}_\mathrm{FRB}$, $\sigma_\mathrm{DM}$ and $\Delta \theta_\mathrm{FRB}$ can affect our cross-correlation measurements in Subsection~\ref{subsubsec:s2n} in details.

\subsection{Statistical errors}\label{subsec:stats_error}

To compute the statistical error, 
we assume that the relevant random fields of $\delta^{\rm 2D}_{\rm h}$ and $D_\mathrm{obs}$ follow Gaussian statistics.
Although either field is not Gaussian in reality, we expect that the statistical error in our cross-correlation analysis is largely determined by the finite sampling effect of FRBs \citep[i.e.~the shot noise in Eq.~(\ref{eq:c_dd}); also see][for an acutual data analysis]{2021ApJ...922...42R}. When the finite sampling effect is dominated, any non-Gaussian contributions to the statistical errors would be less important \citep[e.g.~see Section~VII in][]{2020PhRvD.102b3528R}.

When we measure the cross power spectrum at bins of multipole $\ell$ with 
a bin width being $\Delta \ell$, the covariance matrix of two power spectra is \en{expressed as follows:}
\beqa
\!\!\!\!&&\!\!\!\!
\mathrm{Cov}\left[
C_\mathrm{hD}(\ell_i, z_\mu, M_{\mathrm{thr}, p}), C_\mathrm{hD}(\ell_j, z_\nu, M_{\mathrm{thr}, q})\right] \nonumber \\
\!\!\!\!&=&\!\!\!\! \frac{\delta^\mathrm{K}_{ij}}{f_\mathrm{sky}\, 2 \ell_i\, \Delta \ell}
\Biggl[\delta^\mathrm{K}_{\mu\nu} C_\mathrm{DD}(\ell_i)\, C_\mathrm{hh}(\ell_i; z_\mu, M_{\mathrm{thr}, p}, M_{\mathrm{thr}, q}) \nonumber \\
&&\!\!\!\!
+ C_\mathrm{hD}(\ell_i, z_\mu, M_{\mathrm{thr}, p})\, C_\mathrm{hD}(\ell_i, z_\nu, M_{\mathrm{thr}, q}) \Biggr],
\label{eq:cov}
\eeqa
where $C_\mathrm{hD}(\ell_i, z_\mu, M_{\mathrm{thr}, p})$ is the cross power spectrum at $\ell=\ell_i$
for haloes with $M\ge M_{\mathrm{thr}, p}$ at the $\mu$-th redshift bin,
$C_\mathrm{DD}$ is the auto power spectrum of $D_\mathrm{obs}$,
$C_\mathrm{hh}(\ell; z_\mu, M_{\mathrm{thr}, p}, M_{\mathrm{thr}, q})$
represents the power spectrum between two halo samples of $M\ge M_{\mathrm{thr}, p}$
and $M \ge M_{\mathrm{thr}, q}$ at the $\mu$-th redshift bin,
$f_\mathrm{sky}$ is the fraction of the sky coverage,
and $\delta^\mathrm{K}_{ij}$ is the Kronecker delta symbol.

Using the Limber approximation, we write the auto power spectrum of $D_\mathrm{obs}$ as \en{follows:}
\beqa
C_\mathrm{DD}(\ell) \!\!\!\!&\simeq&\!\!\!\!  C_\mathrm{DD, IGM}(\ell) \, {\cal B}^2(\ell) + \frac{\sigma^2_\mathrm{DM}}{\bar{n}_\mathrm{FRB}}, \label{eq:c_dd} \\
C_\mathrm{DD, IGM}(\ell) \!\!\!\!&=&\!\!\!\! \int_0^{\chi_H}\! \mathrm{d}\chi\, \frac{K^2_\mathrm{e}(\chi)}{\chi^2}\,
P_\mathrm{e}\left(\frac{\ell}{\chi}, z(\chi)\right),
\eeqa
where $P_\mathrm{e}(k,z)$ is the 3D auto power spectrum of $\delta_\mathrm{e}(\bx)$.
We adopt the fitting formula of $P_\mathrm{e}(k,z)$ in \citet{2021MNRAS.502.2615T}.
Note that we ignore contributions from the clustering of sources in Eq.~(\ref{eq:c_dd}), but \en{they} are expected to be sub-dominant \citep[e.g.][]{2017PhRvD..95h3012S}.
Similarly, the power spectrum of $C_\mathrm{hh}$ is \en{expressed as follows:}
\beqa
C_\mathrm{hh}(\ell; z_\mu, M_{\mathrm{thr}, p}, M_{\mathrm{thr}, q})
\!\!\!\!&=&\!\!\!\! C_\mathrm{hh, cl}(\ell; \mu, p, q) 
+\frac{\delta^\mathrm{K}_{pq}}{\bar{n}^{\rm 2D}_\mathrm{h, \mu p}}, \label{eq:c_hh} \\
C_\mathrm{hh, cl}(\ell; \mu, p, q)
\!\!\!\!&=&\!\!\!\!
\int_{\chi_{\mathrm{min},\mu}}^{\chi_{\mathrm{max},\mu}} \!
\mathrm{d}\chi\, \chi^2\, P_\mathrm{L}\left(\frac{\ell}{\chi},z(\chi)\right) \nonumber \\
&&\!\!\!\!
\times \,
b_\mathrm{S,\mu p}(z(\chi))\, b_\mathrm{S,\mu q}(z(\chi))
\eeqa
where $\chi_{\mathrm{min},\mu}$ and $\chi_{\mathrm{max},\mu}$ set 
the boundary in the range of comoving distances at the $\mu$-th redshift bin,
and 
\beqa
\bar{n}^{\rm 2D}_\mathrm{h, \mu p} \!\!\!\!&=&\!\!\!\!
\int_{\chi_{\mathrm{min}, \mu}}^{\chi_{\mathrm{max}, \mu}} \! \mathrm{d}\chi\, \chi^2
\int \!\mathrm{d}M \frac{\mathrm{d}n}{\mathrm{d}M}(M,\chi) \, S_\mathrm{p}(M), \\
b_\mathrm{S,\mu p}(z) \!\!\!\!&=&\!\!\!\!
\frac{1}{\bar{n}^{\rm 2D}_\mathrm{h,\mu p}}\int \! \mathrm{d}M\frac{\mathrm{d}n}{\mathrm{d}M}(M,z)\, S_\mathrm{p}(M)\, b_\mathrm{h}(M,z), \\
S_\mathrm{p}(M) \!\!\!\!&\equiv&\!\!\!\! \Theta(M-M_{\mathrm{thr},p}).
\eeqa

\subsection{Fisher matrix}\label{subsec:Fisher}

To quantify information contents in $C_\mathrm{hD}$, we measure the constraining power of some physical parameters based on a Fisher formalism.
Suppose that we measure the cross power spectra $C_\mathrm{hD}$
for three mass-limited halo catalogues at five redshift bins.
Then, we construct the data vector of $\bd{D}$ as \en{follows:}
\beqa
\bd{C}_{\mu, p} \!\!\!\!&=&\!\!\!\!
\{C_\mathrm{hD}(\ell_1, z_\mu, M_{\mathrm{thr}, p}), 
C_\mathrm{hD}(\ell_2, z_\mu, M_{\mathrm{thr}, p}), \nonumber \\
&&
\quad \quad \quad \quad \quad \quad \quad
\cdots, C_\mathrm{hD}(\ell_N, z_\mu, M_{\mathrm{thr}, p})\}
\\
\bd{D} \!\!\!\!&=&\!\!\!\! \{ \bd{C}_{1, 1}, \bd{C}_{1, 2}, \bd{C}_{1, 3},
\bd{C}_{2, 1}, \bd{C}_{2, 2}, \bd{C}_{2, 3}, \nonumber \\
&&
\quad \quad \quad \quad
\quad \quad \quad \quad \quad
\cdots,
\bd{C}_{5, 1}, \bd{C}_{5, 2}, \bd{C}_{5, 3}\}, \label{eq:Dvec}
\eeqa
where we set the number of $\ell$ bins to $N$, and the data vector \en{comprises} $N\times3\times5$ power spectra.
We perform a logarithmic binning in a range of $10<\ell<\ell_\mathrm{max}$ with the logarithmic bin width being $\Delta \ln \ell = 0.4$.
We will examine two cases of $\ell_\mathrm{max}=3000$ and $10000$.
We consider the Fisher analysis with $\ell_\mathrm{max}=3000$ as our baseline, \en{whereas} the results with $\ell_\mathrm{max}=10000$
can give some sense of how much information remains at sub-arcmin scales.

For physical parameters of interest $\bd{s}$, the Fisher matrix is then defined as \en{follows:}
\beqa
F_{ij} &=& 
\sum_{\alpha, \beta}\frac{\partial D_{\alpha}}{\partial s_i}
\, \mathrm{Cov}^{-1}_{\alpha \beta} 
\frac{\partial D_{\beta}}{\partial s_j} \nonumber \\
&& 
+ \frac{1}{2}\sum_{\alpha,\beta,\gamma,\delta}\mathrm{Cov}^{-1}_{\alpha \beta} \frac{\partial \mathrm{Cov}_{\beta \gamma}}{\partial s_i} \mathrm{Cov}_{\gamma \delta}^{-1} \frac{\partial \mathrm{Cov}_{\delta \alpha}}{\partial s_j}, \label{eq:Fisher}
\eeqa
where $\mathrm{Cov}$ is the covariance matrix for the data vector $\bd{D}$, \en{which} is given by Eq.~(\ref{eq:cov}).
Throughout this paper, we ignore the second term in the right-hand side of Eq.~(\ref{eq:Fisher}) because the covariance scales with the inverse of \en{the} survey area, and it is expected to be subdominant for the analysis with $20000 \, {\rm deg}^2$.
When $\Delta s_i$ represents 
the statistical error in the parameter of $s_{i}$ inferred from a 
measurement of $\bd{D}$, the inverse of the Fisher matrix $F^{-1}_{ij}$ 
provides an estimate of the correlation 
between $\Delta s_i$ and $\Delta s_j$.

\begin{table}
 \caption{
 The list of parameters in our Fisher analysis. 
 Assuming three different mass-limited samples, we have 16 parameters in total, consisting of 6 cosmological parameters, 4 astrophysical parameters 
 (see Table~\ref{tab:model_summary} for details),
 2 nuisance parameters in the halo concentration (Eq.~[\ref{eq:c_h}])
 and $f_\mathrm{e}$ (Eq.~[\ref{eq:f_e_model}]),
 3 thresholds of the mass-limited samples,
 and 1 for the FRB redshift distribution (see Eq.~[\ref{eq:pz}]).
 Note that the parameters of $M_c$, $M_{1,0}$ and 
 $M_{\mathrm{thr},i}$ ($i=1,2,3$) are in units of $h^{-1}\,M_\odot$.
 For the fiducial astrophysical parameters, 
 we use the values relevant to the TNG simulation.
 }
 \label{tab:Fisher_summary}
 \begin{center}
 \begin{tabular}{@{}lll}
  \hline
  Parameters & Fiducial & Gaussian prior $\sigma_\mathrm{prior,i}$\\
  \hline
  Cosmology & & \\
  \hline
  $h$ & 0.673 & 0.1 \\
  $w_0$ & -1 & 0.2 \\
  $\Omega_\mathrm{M}$ & 0.315 & 0.10 \\
  $\Omega_\mathrm{b}$ & 0.0491 & 0.020 \\
  $\sigma_8$ & 0.831 & 0.1 \\
  $n_s$ & 0.965 & 0.1 \\
  \hline
  Astrophysics (Gas) & & \\
  \hline
  $\log \eta$ & $\log(0.14)$ & No prior adopted \\
  $\log M_c $ & $\log(2.3\times10^{13})$ & No prior adopted \\
  $\log \beta$ & $\log(4.09)$ & No prior adopted \\
  $\log M_{1,0}$ & $\log(2.2\times10^{10})$ & 0.5 dex \\
  \hline
  Astrophysics (Nuisance) & & \\
  \hline
  $A_\mathrm{DK15}$ & 1 & 0.3 \\
  $A_\mathrm{e,norm}$ & 1 & 0.2 \\
  \hline
  Halo mass & & \\
  \hline
  $\log M_\mathrm{thr,1}$ & 12 & 0.3 dex \\
  $\log M_\mathrm{thr,2}$ & 13 & 0.3 dex \\
  $\log M_\mathrm{thr,3}$ & 14 & 0.3 dex \\
  \hline
  FRB redshifts & & \\
  \hline
  $\alpha$ & 3.5 & No prior adopted \\
  \hline
 \end{tabular}
 \end{center}
\end{table}

In our setup, we have six parameters to compute the linear matter power spectrum $P_\mathrm{L}(k,z)$ \en{and} the property of dark matter haloes: 
the dimensionless Hubble parameter $h$, equation-of-state parameter of the DE $w_0$, matter density $\Omega_\mathrm{M}$, baryon density $\Omega_\mathrm{b}$, spectral index of primordial power spectrum $n_s$ and linear mass variance smoothed with the scale of $8\, h^{-1}\mathrm{Mpc}$ at $z=0$, $\sigma_8$.

In addition, there are four parameters to determine the gas properties around dark matter haloes, referred to as $\eta$, $M_c$, $\beta$ and $M_{1,0}$ (see Table~\ref{tab:model_summary} and \en{Subsection}~\ref{subsec:gas_model} for details) \en{and} two nuisance parameters of $A_\mathrm{DK15}$ and $A_\mathrm{e,norm}$. 
To make our analysis realistic as possible, we allow the mass thresholds of $M_\mathrm{thr} = 10^{12}$, $10^{13}$ and $10^{14}\, h^{-1}M_\odot$ and the parameter of the FRB redshift distribution (i.e. $\alpha$ in Eq.~[\ref{eq:pz}]) to \en{vary}.

Therefore, we have $6+4+2+4 = 16$ free parameters in the model of our data vector $\bd{D}$.
Owing to the strong degeneracy among parameters in the power spectra, 
we need to introduce loose priors for \en{a} realistic forecast of parameter constraints. 
\en{Thus}, we compute the Fisher matrix as follows:
\beq
\bd{F}=\bd{F}_\mathrm{model} + \bd{F}_\mathrm{prior},
\eeq
where $\bd{F}_\mathrm{model}$ is computed \en{from} Eq.~(\ref{eq:Fisher}) with the power spectra of $C_\mathrm{hD}$, and $\bd{F}_\mathrm{prior}$ represents the prior information. 
We assume that the loose prior on the $i$-th parameter is given by Gaussian with the variance, $\sigma_{\mathrm{prior},i}$. 
In this case, the prior term is expressed as $F_{\mathrm{prior},ij}=\delta^\mathrm{K}_{ij}\, \sigma^{-2}_{\mathrm{prior},i}$.

In this paper, we examine if our analysis can bring independent cosmological information from other probes. Hence, we set the prior information of cosmological parameters to be non-informative as possible.
For the cosmological parameters, our prior width is larger than the $10\sigma$ level in the latest analysis of cosmic microwave backgrounds by the Planck \citep[][]{2020A&A...641A...6P}.
Among six cosmological parameters, the prior in $\Omega_\mathrm{b}$ and $h$ is the most important for our cross-correlation analysis. Note that the amplitude of our cross-correlation scales with $\Omega_\mathrm{b}h$.

Table~\ref{tab:Fisher_summary} summarises the fiducial value of the parameter to compute $\bd{F}_\mathrm{model}$ \en{and} the loose prior width of $\sigma_{\mathrm{prior},i}$.
\en{Notably}, the mass selection function (or $M_\mathrm{thr}$ in our case) and the stellar-to-halo mass relation ($M_{1,0}$) can be tightly constrained from other methods, such as 
stacked weak lensing \citep[e.g.][]{2013MNRAS.435.2345H,
2015MNRAS.454.1161Z, 2018ApJ...854..120M} and abundance matching analyses \citep[e.g.][]{2013ApJ...770...57B, 2019MNRAS.488.3143B}.
We also include the prior information of $A_\mathrm{e,norm}$, which is broadly consistent with the analysis in \citet{2020MNRAS.496L..28L}
who provided an observational constraint on $f_{\rm e}$ using the DM-$z$ relation from five localised FRBs.

\section{Results}\label{sec:results}

\subsection{Comparisons \en{of} our model and TNG}\label{subsec:comp_model_TNG}

\begin{figure}
 \includegraphics[width=\columnwidth]{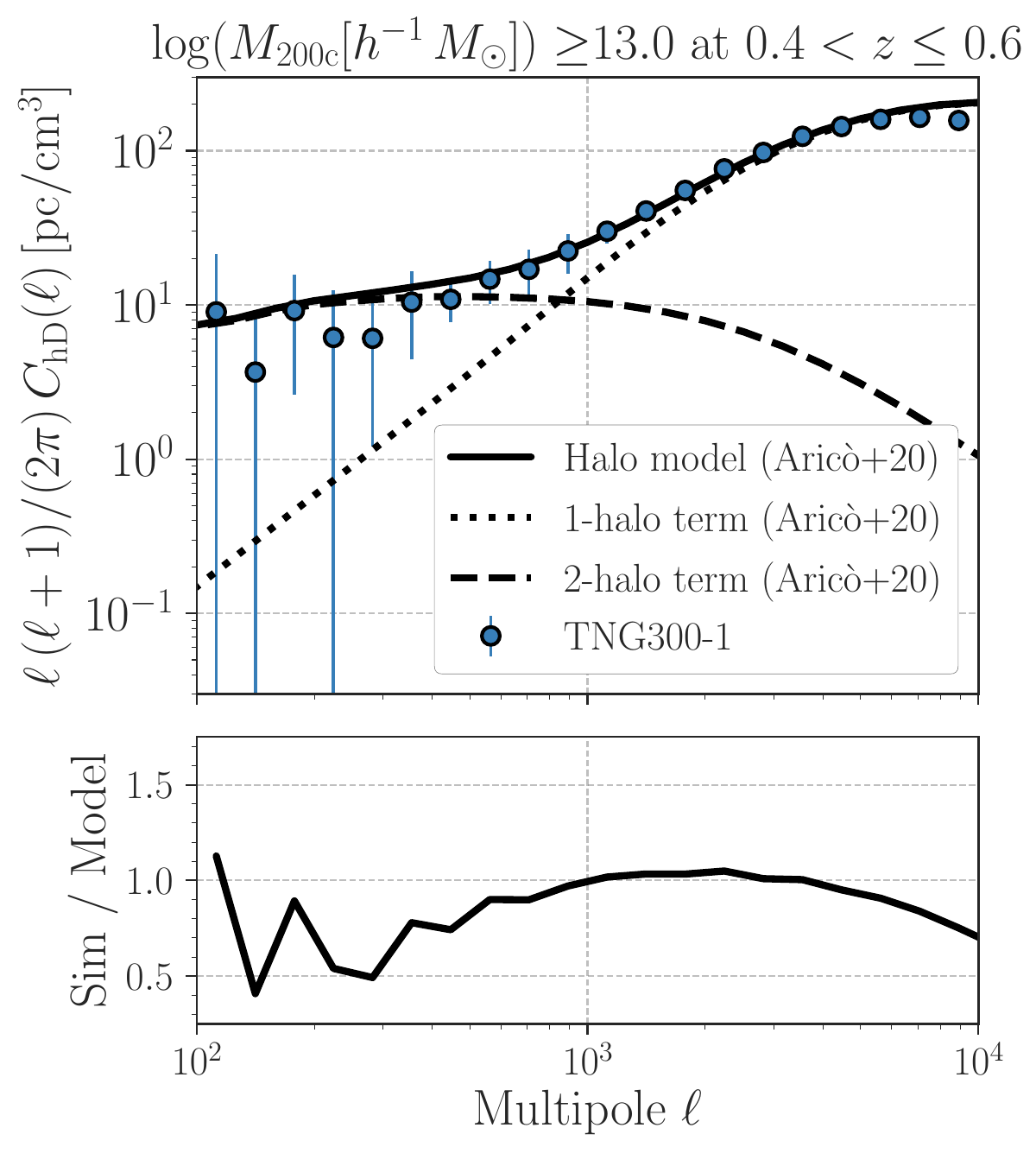}
 \caption{Comparison of our fiducial model of 
 the cross power spectrum of cosmic DM and dark matter haloes (Eq.~[\ref{eq:Cl_limber}] with Eq.~[\ref{eq:halo_model}]) with the counterpart measured in the TNG simulation. 
 We here consider the mass-limited halo sample with $M \ge 10^{13}\, h^{-1}M_\odot$ at the redshift of $0.4<z\le 0.6$ and assume that all the FRBs locate at $z_s = 1$ for simplicity.
 The selected haloes mimic a large sample of massive galaxies observed in 
 the SDSS-III's Baryon Oscillation Spectroscopic Survey \citep{2013MNRAS.432..743N}.
 The blue circles with error bars in the upper panel show the power spectrum in the simulation, while
 the solid line represents our fiducial model. 
 The dotted and dashed lines stand for 1-halo and 2-halo terms, respectively. In the upper panel, the error bars are the standard deviations over 27 realisations of the TNG-based light-cone catalogues \citep{2021MNRAS.502.2615T}. 
 The error bars simply scale as $[({\rm survey~area})/(36 \, {\rm deg}^2)]^{-1/2}$,
 but the source number density is set to $900^2 \, {\rm deg}^{-2} =810,000\, \mathrm{deg}^{-2}$ (the size of error bars is unrealistically small for a fixed survey area).
 The bottom panel shows the ratio of the simulated power spectrum with our model.
 }
 \label{fig:TNG_comp_cls_example}
\end{figure}

We first compare our fiducial model of the cross power spectrum $C_\mathrm{hD}$ with the counterpart in the TNG300-1 simulation \citep{2018MNRAS.480.5113M, 2018MNRAS.477.1206N, 2018MNRAS.475..624N, 2018MNRAS.475..648P, 2018MNRAS.475..676S}.
The TNG team computed the astrophysical processes (such as gas dynamics, star formation and AGN feedback) \en{and} the gravitational evolution using the moving-mesh code {\tt AREPO} \citep{2010MNRAS.401..791S}. 
The box size \en{was} $205 \, h^{-1} {\rm Mpc}$ with the number of particles being $N_{\rm p}=2500^3$, where $N_{\rm p}$ is the same for baryon and dark matter particles at the starting redshift ($z=127$).\footnote{At the initial redshift, there are $2500^3$ gas particles in the simulation, while no stars and black holes exist. As time evolves, star formation begins in high-density regions, where some gas particles are destructed and stellar or black-hole particles are produced with certain conditions. This process does not exactly conserve the number of baryon particles, and thus $N_{\rm p}$ for baryons slightly decreases from the initial number ($2500^3$).}
Each baryon particle has one of three forms: gas, star or black hole.
Free electrons are included in the gas particles.

Mock DM maps have been computed from the projection of gas particles in the TNG simulations at different redshifts in a light cone.
We use 27 realisations\footnote{In the previous work \citep{2021MNRAS.502.2615T}, $10$ DM maps were prepared. In this paper, we added $17$ mock maps repeating the same procedure.} of the light-cone catalogues of dark matter haloes \en{and} DM maps in \citet{2021MNRAS.502.2615T}.
Single light-cone halo catalogues and DM maps cover a sky coverage of $6\times6\, \mathrm{deg}^2$ in which $5400^2$ light rays are homogeneously emitted from the observer; i.e. the resulting angular resolution is $4$ arcsec ($=6 \, {\rm deg}/5400$).
To increase the number of realisations, the simulation coordinates were randomly shifted and swapped when defining the light cone in their ray-tracing pipeline.
We select dark matter haloes by imposing $M_{200\mathrm{c}} \ge 10^{13}\, h^{-1}M_\odot$ 
and $0.4<z\le 0.6$ in the light-cone catalogues and then perform the cross-correlation analysis with DM maps at the source redshift of $z_s=1$.
\en{Notably}, the simulated DM maps do not contain the contribution from $D_\mathrm{MW}$.
We set $\Delta \theta_\mathrm{FRB}=0$ in the cross-correlation analysis for simplicity.
We summarise how to measure the power spectrum from the simulation data and results for halo samples as varying the selection of halo masses and redshifts in Appendix~\ref{apdx:comp_TNG}.
Caution that we set $z_s=1$ only when comparing our model with the simulation results in this subsection and Appendix~\ref{apdx:comp_TNG}. Otherwise, our model adopts the redshift distribution of FRBs in Eq.~(\ref{eq:pz}).

\begin{figure*}
 \includegraphics[width=2\columnwidth]{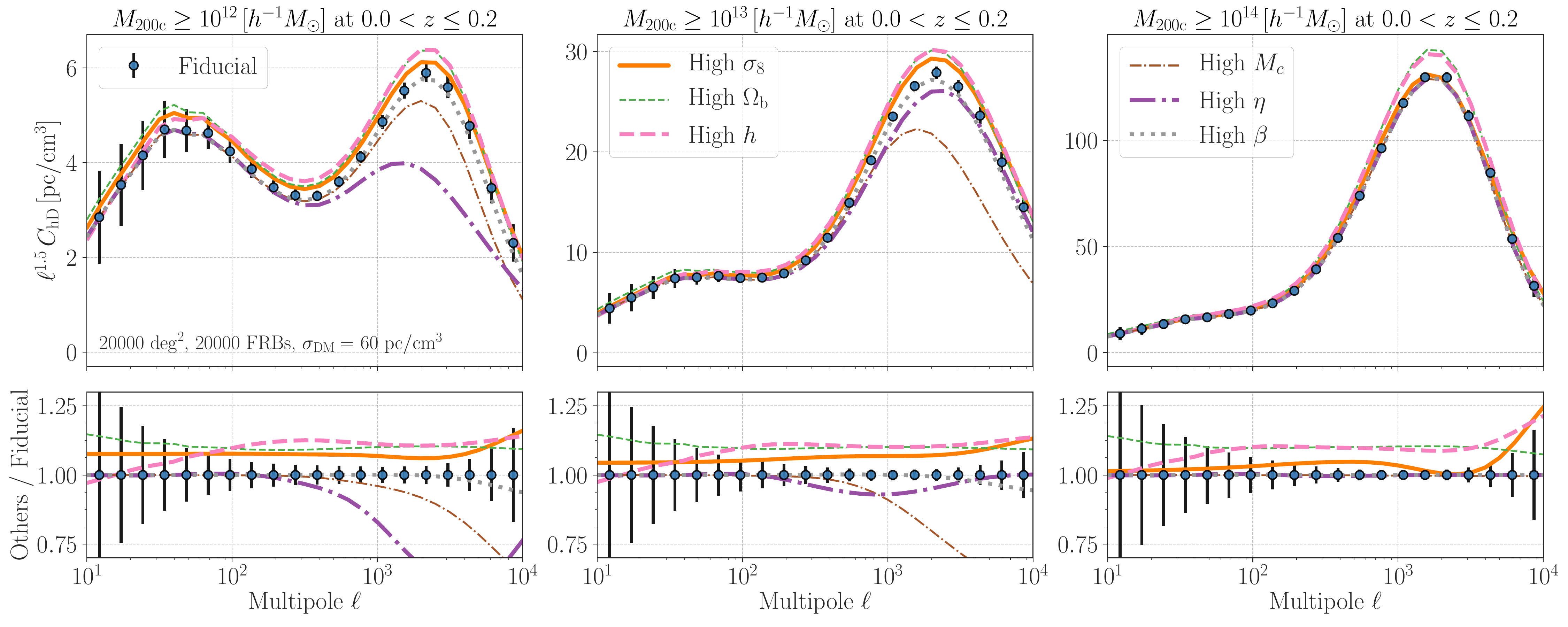}
 \caption{The parameter dependence of cross power spectra between haloes
 and DM.
 The left, middle and right panels show the power spectrum for the mass-limited samples 
 with $M \ge 10^{12}$, $10^{13}$ and $10^{14}\, h^{-1}M_\odot$ at $0<z\le 0.2$, respectively.
 Each top panel shows the change of the power spectra as varying the cosmological parameters of $\Omega_\mathrm{b}$, $h$ and $\sigma_8$, and the astrophysical parameters of $M_c$, $\eta$ and $\beta$
 (see Table~\ref{tab:model_summary} and \en{Subsection}~\ref{subsec:gas_model} for details about our model of gas).
 In the bottom, we show the fractional difference among the power spectra.
 The error bars in this figure is set by the Gaussian statistical uncertainty for the survey sky coverage of $20000 \, {\rm deg}^2$, the mean FRB number density of \en{$1\, \mathrm{deg^{-2}}$} and the variance of host-galaxy DM being $60\, \mathrm{pc}/\mathrm{cm}^3$.
 We here do not include the smearing effect due to localisation errors of FRBs for simplicity.
 The blue points represent our fiducial model with the best-fit cosmological model by \citet{2016A&A...594A..13P}
 and the astrophysics mimicking the TNG simulation \citep{2018MNRAS.475..676S}, 
 while the orange solid, green dotted, pink dashed, brown dashed-dotted, purple dashed-dotted and grey dotted lines are the models with higher $\sigma_8$, $\Omega_\mathrm{b}$, $h$, $M_c$, $\eta$ and $\beta$, respectively.
 Note that $M_c$ gives a halo mass scale above which more than half of gas in the halo is bound, 
 $\eta$ determines a typical distance scale of ejected gas from the halo centre due to feedback processes,
 and $\beta$ controls the gas-to-halo mass relation.
 }
 \label{fig:mass_limited_cls_param}
\end{figure*}

Figure~\ref{fig:TNG_comp_cls_example} shows our baseline model of the cross power spectrum and comparisons with the simulation results.
The dashed line in the upper panel is the 2-halo term arising from the two-point correlation between separated haloes (Eq.~[\ref{eq:P2h}]), \en{whereas} the solid line includes the correlation sourced by free electrons inside single haloes (Eq.~[\ref{eq:halo_model}]).
At $\ell \simgt 10^{3}$ (corresponding angular scale of $\simlt 3\, \mathrm{arcmin}$), the 2-halo term becomes subdominant, and most cross-correlation signals can be explained by free electrons inside haloes.
We find that our model provides a reasonable fit to the simulation result for $\ell\simlt10^{4}$, 
although the model of free electrons is not calibrated with electron density profiles 
in the hydrodynamical simulation \citep[see,][for details of the calibration]{2020MNRAS.495.4800A}.

Note that our model is phenomenological and relies on various fitting formulas for large-scale structures.
To further improve the model precision, we may need to \en{consider} various points, such as possible changes of the polytropic index of gas density at central regions of haloes \citep[see e.g.][for examples of cluster-sized haloes]{2013ApJ...774...23M},
non-linear halo bias \citep[e.g.][]{2021MNRAS.503.3095M}
and baryonic effects on statistical properties of haloes \citep[e.g.][]{2016MNRAS.456.2361B, 2021arXiv210305076B}.
These points should be \en{considered} in the near future; \en{they are} beyond the scope of this \en{study}.
We revisit the limitation of our model in \en{Subsection}~\ref{subsec:model_accuracy}.

\subsection{Information contents}

\subsubsection{Parameter dependence}\label{subsubsec:param_dependence}

Next, we study the dependence of the cross power spectrum on model parameters.
Figure \ref{fig:mass_limited_cls_param} shows some representative examples of predicted power spectra 
as varying relevant cosmological and astrophysical parameters.
In this figure, we consider seven cases as follows:
\begin{description}
\item[(i)] Fiducial cosmological and astrophysical parameters as listed in Table~\ref{tab:Fisher_summary};
\item[(c-i)] The cosmology with a higher $\sigma_8$ by 0.07 and the fiducial astrophysics;
\item[(c-ii)] The cosmology with a higher $\Omega_\mathrm{b}$ by 10\% and the fiducial astrophysics;
\item[(c-iii)] The cosmology with a higher $h$ by 10\% and the fiducial astrophysics;
\item[(a-i)] The fiducial cosmology and the astrophysical model with a higher $M_c$ by 0.2 dex ($=58.4\%$);
\item[(a-ii)] The fiducial cosmology and the astrophysical model with a higher $\eta$ by 0.2 dex;
\item[(a-iii)] The fiducial cosmology and the astrophysical model with a higher $\beta$ by 0.2 dex.
\end{description}
We \en{now} consider the mass-limited samples of 
$M_{200\mathrm{c}} \ge 10^{12}$, $10^{13}$ and $10^{14}\, h^{-1}M_\odot$ at the redshift of $0<z\le 0.2$.
We adopt Eq.~(\ref{eq:pz}) for the redshift distribution of FRBs from this section.
In each panel, the blue circles with error bars show the model (i),
\en{whereas} the orange solid, green dotted, pink dashed, brown dashed-dotted, purple dashed-dotted and grey dotted lines
represent model (c-i), (c-ii), (c-iii), (a-i), (a-ii) and (a-iii), respectively.
The black error bars are estimated as in \en{Subsection}~\ref{subsec:stats_error} when the sky coverage of $20000 \, {\rm deg^2}$, $\bar{n}_\mathrm{FRB}=1\, \mathrm{deg}^{-2}$
and $\sigma_\mathrm{DM}=60\, \mathrm{pc}/\mathrm{cm}^3$ are adopted.
In the figure, we ignore the smearing effect ${\cal B}(\ell)$ for simplicity.
Note that the cross power spectrum will be exponentially suppressed 
at $\ell \simgt 3800\, (\Delta\theta_\mathrm{FRB}/3 \, \mathrm{arcmin})^{-1}$ if the smearing effect is included.
In each panel in the figure, the bump in $\ell^{1.5}\, C_\mathrm{hD}$ at $\ell \simlt 100$ is determined by the 2-halo term, \en{whereas} the counterpart at $\ell \simgt 1000$ is dominated by the 1-halo term.

\begin{figure*}
 \includegraphics[width=2\columnwidth]{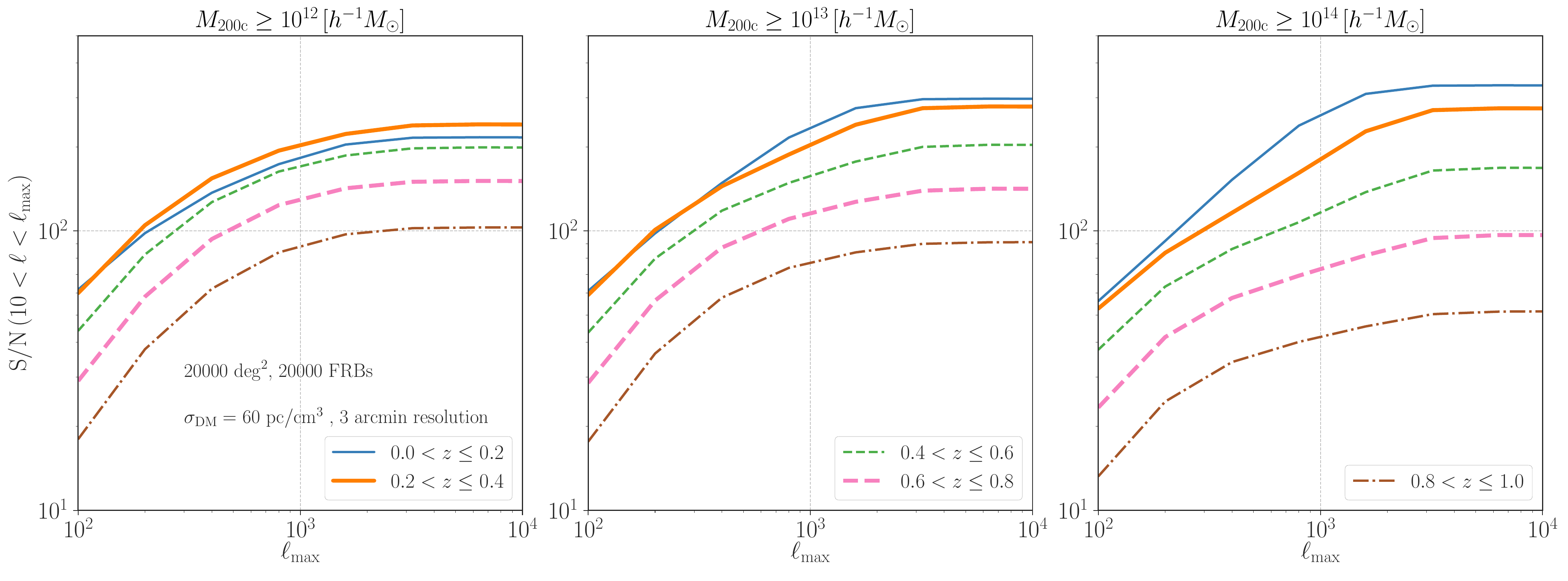}
 \caption{The cumulative SNR (S/N) of cross power spectra between DM and dark matter haloes.
 In each panel, we show the SNR for a given mass-limited halo sample at different redshift bins as varying the maximum multipoles ($\ell_\mathrm{max}$) used in the cross-correlation analysis.
 The blue thin, orange thick, green dashed, pink think dashed and brown dotted-dashed lines show the SNR at the redshift bin of $0<z\le0.2$, $0.2<z\le0.4$, $0.4<z\le0.6$, $0.6<z\le0.8$ and $0.8<z\le 1$, respectively.
 The left, middle and right panels represent the results for the mass-limited samples with $M\, [h^{-1}M_\odot]\ge 10^{12}$, $10^{13}$ and $10^{14}$, respectively.
 In this figure, we assume that 20000 FRBs are available over $20000 \, {\rm deg}^2$ and the variance of host-galaxy DM being
 $60\, \mathrm{pc}/\mathrm{cm}^3$ with the localisation error of 3 arcmin.
 Note that a measurement with $\mathrm{S/N}=100$ indicates that 
 the amplitude of the cross power spectrum can be measured with a 1\% level of precision.
 }
 \label{fig:mass_limited_s2n_eachz_eachm}
\end{figure*}

\subsubsection*{Cosmological dependence}

Models (c-i)--(c-iii) highlight the cosmological dependence on the cross power spectrum.
A higher $\Omega_\mathrm{b}$ can increase the abundance of free electrons in the universe,
enhancing the cross power spectra regardless of the halo mass threshold.
\en{A} similar enhancement can be seen when one increases the Hubble parameter $h$ because the overall amplitude of cosmic \en{DM} scales with $\Omega_\mathrm{b}h$ (see Eqs.~[\ref{eq:D_LSS_singlez}] and [\ref{eq:W_e}]).

Nevertheless, the cross-correlation allows us to break 
the degeneracy of $\Omega_\mathrm{b}$ and $h$ using the 2-halo term.
The 2-halo term is proportional to the linear matter power spectrum $P_\mathrm{L}$, and the location of the peak of $P_\mathrm{L}$ is determined by the scale of matter-radiation equality.
The comoving wave number at the peak relates with the horizon scale at the matter-radiation equality and varies with $\Omega_\mathrm{M}h^2$ \citep[e.g.][]{1998ApJ...496..605E}.
Hence, increasing $h$ can make the 2-halo term of $C_\mathrm{hD}$ \en{shift} toward higher $\ell$.
This effect causes the difference of the 2-halo term between models (c-ii) and (c-iii).
The dependence of the 1-halo term on $\Omega_\mathrm{b}$ and $h$ is also helpful to break the degeneracy because the halo concentration can depend on the shape of $P_\mathrm{L}$, which is determined by the physical baryon density $\Omega_\mathrm{b}h^2$.
\en{Notably}, the parameters of $\Omega_\mathrm{b}$ and $h$ can change the shape of $C_\mathrm{hD}$ compared \en{with} the fiducial case, whereas the fraction of electrons in cosmic baryon density $f_\mathrm{e}$ changes the amplitude of $C_\mathrm{hD}$ alone.
Hence, a detailed cross-correlation analysis enables us to break the degeneracy among $\Omega_\mathrm{b}$, $h$ and $f_\mathrm{e}$.
Note that the cross-correlation with galaxies at different redshifts can provide information of the redshift dependence of $f_\mathrm{e}$ in principle.

Because the halo concentration $c_\mathrm{h}$ increases with $\sigma_8$ \citep[e.g.][]{2014MNRAS.441..378L}, the gravitational potential of a halo deepens, and baryons can be more concentrated in our model.
This effect makes the 1-halo term of $C_\mathrm{hD}$ higher as $\sigma_8$ \en{increases}.
On the other hand, the $\sigma_8$ dependence of the 2-halo term can change \en{with a different} mass threshold.
The 2-halo term is proportional to $P_\mathrm{L}$ and the linear halo bias $b_\mathrm{h}$.
The halo bias for more massive haloes decreases with $\sigma_8$, but the bias of galaxy-sized haloes is 
less sensitive to $\sigma_8$ \citep[e.g.][]{1996MNRAS.282..347M, 1999MNRAS.308..119S}.
This effect makes the 2-halo term larger at lower mass-limited samples.
Hence, the degeneracy between $\Omega_\mathrm{b}h$ and $\sigma_8$ can be efficiently broken 
with multiple mass-limited samples.

For other cosmological parameters, we briefly summarise prominent effects in the cross-correlation analysis.
The parameter $n_s$ changes the $\ell$ dependence of the 2-halo term at $\ell \simlt 100$ regardless of halo selections, \en{whereas} the change \en{in} the halo concentration by $n_s$ affects the shape of 1-halo term.
The redshift dependence of $C_\mathrm{hD}$ is useful to determine the cosmological parameters of $\Omega_\mathrm{M}$ and $w_0$ because these parameters govern the expansion rate at the late-time universe (Eq.~[\ref{eq:Hz}]).
In general relativity, the expansion rate also determines the redshift dependence of the linear matter power spectrum \citep[e.g.][]{2005PhRvD..72d3529L}.
Hence, the measurement of the 2-halo terms at different redshifts allows us to constrain $\Omega_\mathrm{M}$ and $w_0$.

\begin{figure*}
 \includegraphics[width=2\columnwidth]{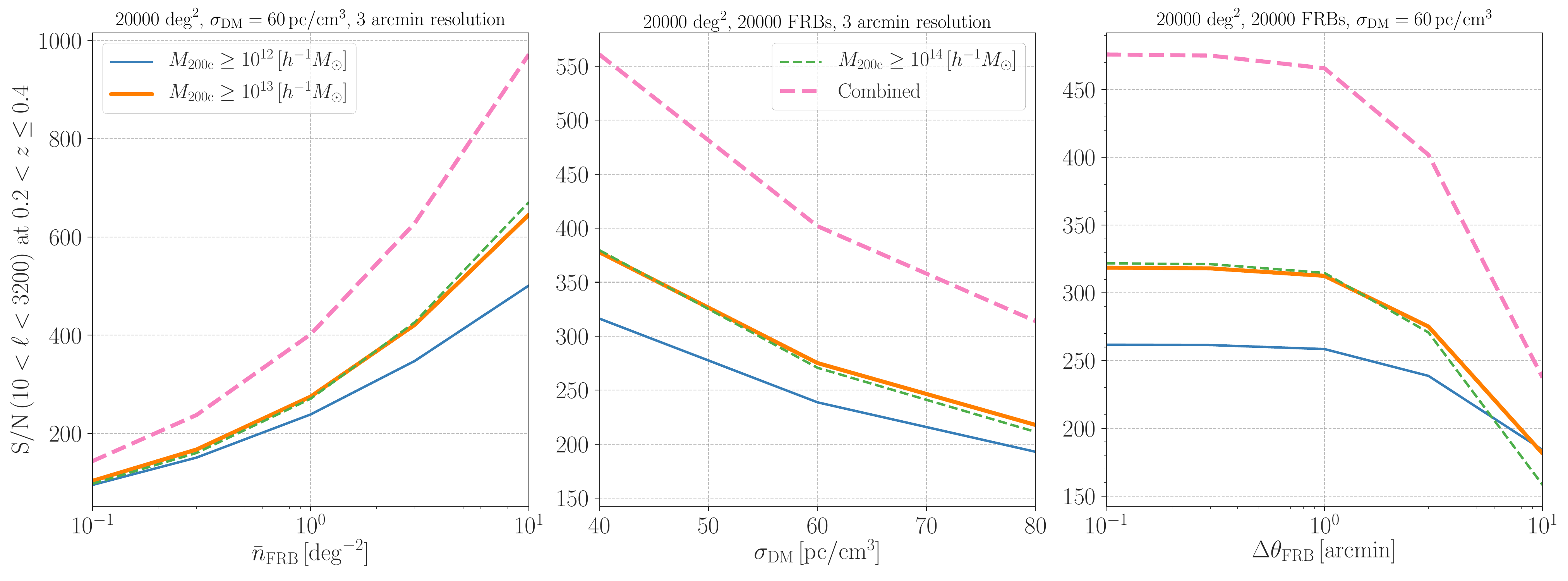}
 \caption{Similar to Figure~\ref{fig:mass_limited_s2n_eachz_eachm}, 
 we vary survey parameters in a hypothetical FRB survey with a sky coverage of $20000 \, {\rm deg}^2$.
 In each panel, we show the SNR for the cross power spectra at $0.2<z\le0.4$ when setting a range of $10 < \ell < 3200$.
 The left panel assumes that FRB catalogues with a variance of host-galaxy DM being $\sigma_\mathrm{DM}=60\, \mathrm{pc}/\mathrm{cm}^3$
 and an angular resolution of FRBs being $\Delta \theta_\mathrm{FRB}=3\, \mathrm{arcmin}$, while we vary the number density of FRBs $\bar{n}_\mathrm{FRB}$.
 The middle panel sets the fiducial values of $\bar{n}_\mathrm{FRB}=1\, \mathrm{deg}^{-2}$ and 
 $\Delta \theta_\mathrm{FRB}=3\, \mathrm{arcmin}$, while we vary $\sigma_\mathrm{DM}$.
 The right panel represents the $\Delta \theta_\mathrm{FRB}$ dependence of the SNR when we set $\bar{n}_\mathrm{FRB}=1\, \mathrm{deg}^{-2}$ and $\sigma_\mathrm{DM}=60\, \mathrm{pc}/\mathrm{cm}^3$.
 The blue thin, orange thick and green dashed lines are the SNRs for the mass-limited samples with $M\, [h^{-1}M_\odot]\ge 10^{12}$, $10^{13}$ and $10^{14}$, while the pink dashed line shows the combined SNR for the three mass-limited samples.
 Note that the combined analysis with different mass-limited samples is still meaningful because
 the cross-correlation signal is mainly determined by a majority of haloes in each sample.}
 \label{fig:mass_limited_s2n_survey_prop}
\end{figure*}

\subsubsection*{Astrophysical dependence}

The difference among models (a-i), (a-ii) and (a-iii) shows the dependence of the cross power spectrum on the astrophysical parameters.
First, we find that cluster-sized haloes are less sensitive to the astrophysical parameters provided the parameters around the TNG-like values \en{are considered}.
Because most baryons around clusters are made of free electrons and bound within their virial regions, the 1-halo term for the sample with $M_{200\mathrm{c}} \ge 10^{14}\, h^{-1}M_\odot$ is 
mainly determined by the amount of cosmic baryons in a cluster, 
scaling with $\sim (\Omega_\mathrm{b}/\Omega_\mathrm{M})\, M_{200\mathrm{c}}$.
As going to lower masses, stars and ejected gas affect the 1-halo term.
Increasing $M_c$ means that the amount of bound gas decreases at $M_{200\mathrm{c}} \simlt 10^{13}\, h^{-1}M_\odot$, making the 1-halo term for $M_\mathrm{200c} \simlt 10^{13}\, h^{-1}M_\odot$ smaller.
When setting $\eta$ larger, the amount of free electrons within a halo decreases at a fixed halo mass.
Because the ejected gas plays a role in the haloes with $M_{200\mathrm{c}} \ll M_c$ in our model, the cross-correlation with a smaller mass $M_{200\mathrm{c}}$ is more sensitive to the change in $\eta$.
The parameter $\beta$ controls how rapidly the gas is expelled from haloes as the halo masses decrease.
A larger $\beta$ increases the amount of bound gas at $M_{200\mathrm{c}} > M_c$, whereas the bound gas decreases at $M_{200\mathrm{c}} < M_c$.
Because our fiducial model assumes $M_c \sim 2\times 10^{13}\, h^{-1}M_\odot$, the cross power spectra for the mass-limited samples of $M_{200\mathrm{c}} \ge 10^{12}$ and $10^{13}\, h^{-1}M_\odot$
are affected by the change in $\beta$.
\en{Notably}, the 2-halo terms are less sensitive to the astrophysical parameters regardless of the halo selection.

\begin{figure}
 \includegraphics[width=1.1\columnwidth, bb = 0 0 547 450 ]{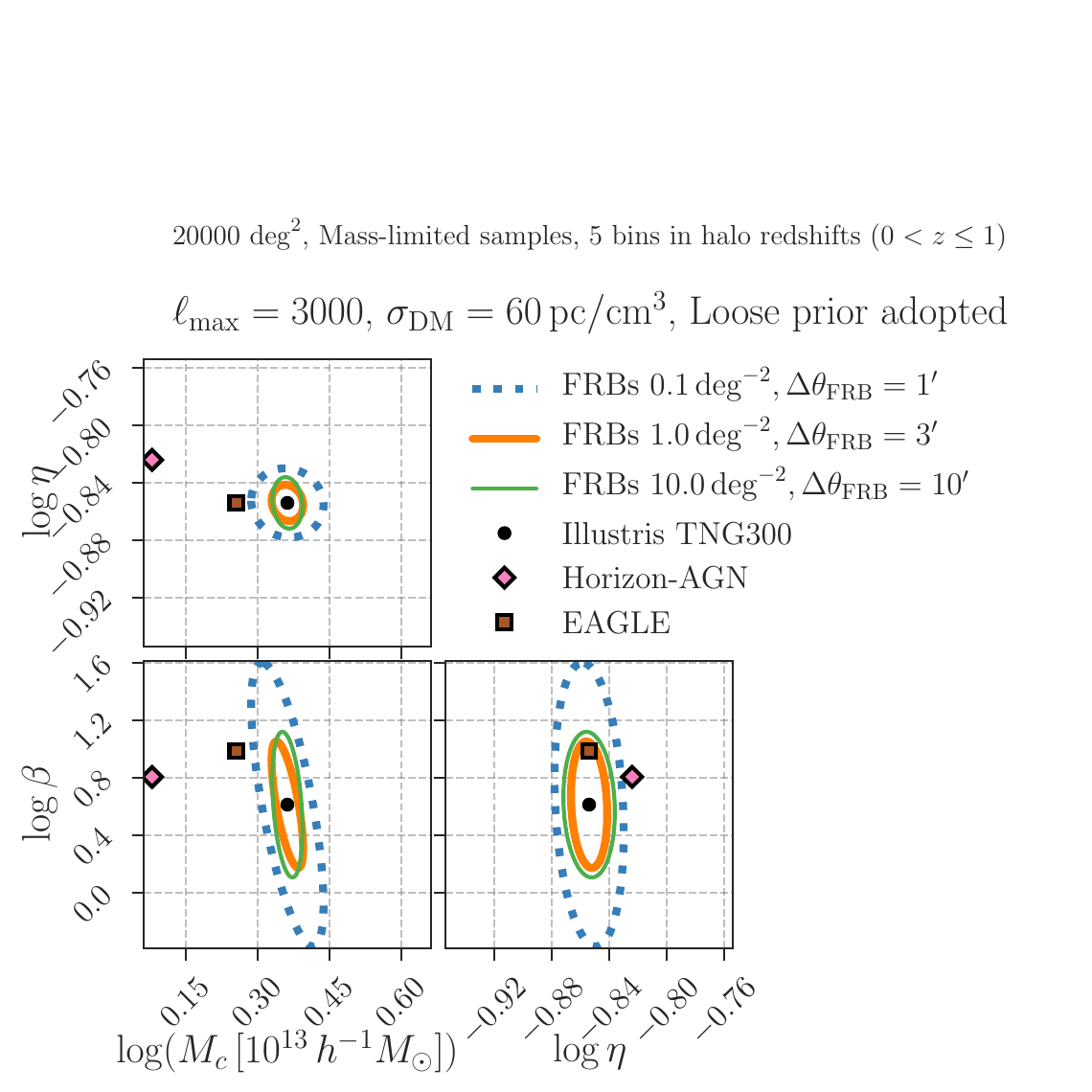}
 \caption{Expected $1\sigma$ constraints of astrophysical parameters about gas in dark matter haloes.
 We consider three possible cases of future FRB data with 
 $(\bar{n}_\mathrm{FRB}, \Delta \theta_\mathrm{FRB}) = (1\, \mathrm{deg}^{-2}, 3')$,
 $(0.1\, \mathrm{deg}^{-2}, 1')$ and $(10\, \mathrm{deg}^{-2}, 10')$,
 where $\bar{n}_\mathrm{FRB}$ is the mean number density of FRBs and 
 $\Delta \theta_\mathrm{FRB}$ is the localisation error.
 The three parameters of $\eta, \beta$, and $M_c$ provide a typical diffusion scale of gas ejected from the halo centre,
 the dependence of the gas mass on the halo mass, and a characteristic mass scale in the gas-to-halo mass relation (see Subsection~\ref{subsec:gas_model} for details).
 We assume that three mass-limited samples at five redshift bins up to $z=1$ are available
 for the cross-correlation analysis. The sky coverage for the analysis is set to $20000 \, {\rm deg}^2$. We use the information about the power spectra in the range of $10<\ell<3000$.
 The black dot shows the best-fit gas model to the TNG simulation results \citep{2018MNRAS.475..676S},
 while the pink diamond and the brown square symbols represent the models suitable for
 the Horizon-AGN \citep{2014MNRAS.444.1453D} and EAGLE simulations \citep{2015MNRAS.446..521S, 2016MNRAS.461L..11H}, respectively. See \citet{2020MNRAS.495.4800A} for details about the calibration of the three parameters.
 }
 \label{fig:mass_limited_gas_params_forecast}
\end{figure}

\subsubsection{Signal-to-noise ratio} \label{subsubsec:s2n}

Another important measure of the information contents is the \en{SNR} (S/N).
It is defined \en{as follows:}
\beq
(\mathrm{S/N})^2 = \sum_{\alpha \beta} D_\alpha \, \mathrm{Cov}^{-1}_{\alpha \beta} D_{\beta}, \label{eq:s2n}
\eeq
where our data vector $\bd{D}$ is given by Eq.~(\ref{eq:Dvec}),
and we define the covariance matrix in Eq.~(\ref{eq:cov}).
In Eq.~(\ref{eq:s2n}), the sum takes over a range of $\ell_\mathrm{min} < \ell < \ell_\mathrm{max}$.
We perform a logarithmic binning in $\ell$ with $\ell_\mathrm{min}=10$ and $\Delta \ln \ell =0.4$.

Figure~\ref{fig:mass_limited_s2n_eachz_eachm} shows the SNR as a function of $\ell_\mathrm{max}$ when we assume that $20000$ FRBs are available over $20000 \, {\rm deg}^2$ and set $\sigma_\mathrm{DM}=60\, \mathrm{pc}/\mathrm{cm}^3$ and the localisation error of 
$\Delta \theta_\mathrm{FRB} = 3\, \mathrm{arcmin}$.
We find that a mass-limited halo sample with a higher mass threshold 
at lower $z$ tends to show a larger SNR.
Cross-correlation analyses with a large $\ell_\mathrm{max}$ make the SNR higher in general, but this is not the case when we include the smearing effect, as in Eq.~(\ref{eq:beam}).
Because we work with $\Delta \theta_\mathrm{FRB} = 3\, \mathrm{arcmin}$ in the figure, the SNR cannot increase at $\ell_\mathrm{max} \simeq \sqrt{16 \ln 2}/\Delta \theta_\mathrm{FRB} \sim 3800$ or larger.
Nevertheless, our halo model predicts that the SNR for most mass-limited samples can reach an order of $100$ if we will be able to use $20000$ FRBs over $20000 \, {\rm deg}^2$.
This indicates that the amplitude of $C_\mathrm{hD}$ can be measured with a $1\%$ level \en{of} precision in future cross-correlation analyses.
The large SNR is mainly driven by the 1-halo terms (the correlation \en{due to} free electrons in 
single haloes).

Figure~\ref{fig:mass_limited_s2n_survey_prop} summarises how the SNR depends on the parameters for a hypothetical FRB survey. In the figure, we consider the cross power spectra of the three mass-limited samples at $0.2<z\le0.4$ in $20000 \, {\rm deg}^2$, whereas we vary one of the survey parameters of the mean number density of FRBs $\bar{n}_\mathrm{FRB}$,
the variance of host-galaxy DM $\sigma_\mathrm{DM}$ and the angular resolution of FRBs $\Delta \theta_\mathrm{FRB}$.
For the cross power spectra at $10<\ell<3200$, the available number of FRBs in a given survey coverage is a primary parameter to determine the SNR.
We find that the SNR can be degraded by a factor of $2$--$3$ when changing $\bar{n}_\mathrm{FRB} = 1\, \mathrm{deg}^{-2}$ to $0.1\, \mathrm{deg}^{-2}$.
The variance of host-galaxy DM $\sigma_\mathrm{DM}$ affects the SNR as well.
The SNR can change by $\sim 1.5$--$2$ within a plausible range of 
$40 \le \sigma_\mathrm{DM} \, [\mathrm{pc}/\mathrm{cm}^3] \le 80$.
To extract full information in the power spectra at $10<\ell<3200$,
we require \en{an} angular resolution of 1 arcmin.
The figure indicates that collecting many FRBs without a detailed localisation is not optimal to have a great SNR at $\ell \simlt 3000$.

\subsection{Fisher forecasts}\label{subsec:Fisher_results}


\begin{table*}
 \caption{Summary of the Fisher forecast of the cosmological and astrophysical parameters by the cross-correlation analysis with DM and dark matter haloes.
 We assume an effective survey area to be $20000 \, {\rm deg}^2$ used for the analysis. 
 We consider three mass-limited samples with $M\ge10^{12}$, $10^{13}$ and $10^{14}\, h^{-1}M_\odot$
 at five separated redshift bins covering up to $z=1$.
 For a hypothetical FRB catalogue, we examine three cases of $(\bar{n}_\mathrm{FRB}, \Delta \theta_\mathrm{FRB})=(1\, \mathrm{deg}^{-2}, 3')$, $(0.1\, \mathrm{deg}^{-2}, 1')$
 and $(10.0\, \mathrm{deg}^{-2}, 10')$,
 where $\bar{n}_\mathrm{FRB}$ is the average number density of FRBs and $\Delta \theta_\mathrm{FRB}$ is the localisation error for each FRB.
 In each table cell, the number without brackets show the $1\sigma$ constraint of single parameter when we marginalise other parameters, while the number in brackets is the marginalised error divided by the fiducial parameter in percentiles.
 We highlight the limit within a 5\% level of precision in black bold, while the red bold font represents the limit better than a 1\% level of precision.
 The details of the Fisher analysis is provided in \en{Subsection}~\ref{subsec:Fisher}. 
 }
 \label{tab:Fisher_forecast_summary}
 \begin{tabular}{lccccc}
  \hline
  Parameters & 
  $\ell_\mathrm{max}=3000$ w/o prior & $\ell_\mathrm{max}=3000$ w/ prior & 
  $\ell_\mathrm{max}=10000$ w/o prior & $\ell_\mathrm{max}=10000$ w/ prior \\
  \hline
  $\bar{n}_\mathrm{FRB}=1\, \mathrm{deg}^{-2}$ and $\Delta \theta_\mathrm{FRB}=3\, \mathrm{arcmin}$ & & & & \\
  \hline
  $h$ & 	 0.0646 ($9.61\%$) & 	0.0354 ($5.26\%$) & 	0.0590 ($8.76\%$) & 	0.0342 ($5.08\%$) \\ 
  $w_0$ & 	 {\bf 0.0422} ($4.22\%$) & 	{\bf 0.0389} ($3.89\%$) & 	{\bf 0.0396} ($3.96\%$) & 	{\bf 0.0371} ($3.71\%$) \\ 
  $\Omega_\mathrm{M}$ & 	 {\bf 0.0126} ($4.00\%$) & 	{\bf 0.0109} ($3.45\%$) & 	{\bf 0.0117} ($3.71\%$) & 	{\bf 0.0102} ($3.23\%$) \\ 
  $\Omega_\mathrm{b}$ & 	 0.0094 ($19.05\%$) & 	0.0057 ($11.64\%$) & 	0.0085 ($17.26\%$) & 	0.0055 ($11.22\%$) \\ 
  $\sigma_8$ & 	 {\bf 0.0118} ($1.42\%$) & 	{\bf 0.0088} ($1.06\%$) & 	{\bf 0.0112} ($1.35\%$) & 	{\bf 0.0085} ($1.03\%$) \\ 
  $n_s$ & 	 {\bf 0.0223} ($2.31\%$) & 	{\bf 0.0181} ($1.88\%$) & 	{\bf 0.0191} ($1.98\%$) & 	{\bf 0.0160} ($1.65\%$) \\ 
  $\log \eta$ & 	 {\bf 0.0129} ($1.51\%$) & 	\bfred{0.0084} ($0.98\%$) & 	{\bf 0.0120} ($1.41\%$) & 	\bfred{0.0079} ($0.93\%$) \\ 
  $\log M_c$ & 	 \bfred{0.0227} ($0.17\%$) & 	\bfred{0.0218} ($0.16\%$) & 	\bfred{0.0212} ($0.16\%$) & 	\bfred{0.0204} ($0.15\%$) \\ 
  $\log \beta$ & 	 0.2973 ($48.60\%$) & 	0.2915 ($47.65\%$) & 	0.2399 ($39.21\%$) & 	0.2365 ($38.66\%$) \\ 
  $A_\mathrm{e,norm}$ & 	 0.2765 ($27.65\%$) & 	0.1425 ($14.25\%$) & 	0.2498 ($24.98\%$) & 	0.1377 ($13.77\%$) \\ 
  $\alpha$ & 	 \bfred{0.0277} ($0.79\%$) & 	\bfred{0.0270} ($0.77\%$) & 	\bfred{0.0270} ($0.77\%$) & 	\bfred{0.0263} ($0.75\%$) \\  
  \hline 
  $\bar{n}_\mathrm{FRB}=0.1\, \mathrm{deg}^{-2}$ and $\Delta \theta_\mathrm{FRB}=1\, \mathrm{arcmin}$ & & & & \\
  \hline
  $h$ & 	 0.1457 ($21.67\%$) & 	0.0483 ($7.19\%$) & 	0.1117 ($16.60\%$) & 	0.0451 ($6.71\%$) \\ 
  $w_0$ & 	 0.1061 ($10.61\%$) & 	0.0885 ($8.85\%$) & 	0.0826 ($8.26\%$) & 	0.0723 ($7.23\%$) \\ 
  $\Omega_\mathrm{M}$ & 	 0.0309 ($9.80\%$) & 	0.0222 ($7.04\%$) & 	0.0237 ($7.53\%$) & 	0.0177 ($5.61\%$) \\ 
  $\Omega_\mathrm{b}$ & 	 0.0191 ($38.85\%$) & 	0.0077 ($15.70\%$) & 	0.0141 ($28.72\%$) & 	0.0072 ($14.69\%$) \\ 
  $\sigma_8$ & 	 {\bf 0.0282} ($3.39\%$) & 	{\bf 0.0171} ($2.06\%$) & 	{\bf 0.0233} ($2.80\%$) & 	{\bf 0.0145} ($1.74\%$) \\ 
  $n_s$ & 	 0.0529 ($5.49\%$) & 	{\bf 0.0351} ($3.64\%$) & 	{\bf 0.0304} ($3.15\%$) & 	{\bf 0.0212} ($2.20\%$) \\ 
  $\log \eta$ & 	 {\bf 0.0297} ($3.47\%$) & 	{\bf 0.0159} ($1.87\%$) & 	{\bf 0.0227} ($2.66\%$) & 	{\bf 0.0113} ($1.32\%$) \\ 
  $\log M_c$ & 	 \bfred{0.0535} ($0.40\%$) & 	\bfred{0.0500} ($0.37\%$) & 	\bfred{0.0365} ($0.27\%$) & 	\bfred{0.0353} ($0.26\%$) \\ 
  $\log \beta$ & 	 0.6753 ($110.39\%$) & 	0.6540 ($106.92\%$) & 	0.3248 ($53.09\%$) & 	0.3177 ($51.94\%$) \\ 
  $A_\mathrm{e,norm}$ & 	 0.5906 ($59.06\%$) & 	0.1644 ($16.44\%$) & 	0.4377 ($43.77\%$) & 	0.1575 ($15.75\%$) \\ 
  $\alpha$ & 	 {\bf 0.0681} ($1.95\%$) & 	{\bf 0.0630} ($1.80\%$) & 	{\bf 0.0591} ($1.69\%$) & 	{\bf 0.0544} ($1.55\%$) \\
  \hline
  $\bar{n}_\mathrm{FRB}=10\, \mathrm{deg}^{-2}$ and $\Delta \theta_\mathrm{FRB}=10\, \mathrm{arcmin}$ & & & & \\
  \hline
  $h$ & 	 0.0599 ($8.90\%$) & 	{\bf 0.0335} ($4.98\%$) & 	0.0598 ($8.89\%$) & 	{\bf 0.0334} ($4.97\%$) \\ 
  $w_0$ & 	 {\bf 0.0427} ($4.27\%$) & 	{\bf 0.0400} ($4.00\%$) & 	{\bf 0.0424} ($4.24\%$) & 	{\bf 0.0398} ($3.98\%$) \\ 
  $\Omega_\mathrm{M}$ & 	 {\bf 0.0130} ($4.13\%$) & 	{\bf 0.0105} ($3.33\%$) & 	{\bf 0.0129} ($4.08\%$) & 	{\bf 0.0104} ($3.30\%$) \\ 
  $\Omega_\mathrm{b}$ & 	 0.0081 ($16.54\%$) & 	0.0051 ($10.46\%$) & 	0.0082 ($16.60\%$) & 	0.0051 ($10.43\%$) \\ 
  $\sigma_8$ & 	 {\bf 0.0125} ($1.51\%$) & 	{\bf 0.0088} ($1.06\%$) & 	{\bf 0.0125} ($1.50\%$) & 	{\bf 0.0088} ($1.06\%$) \\ 
  $n_s$ & 	 {\bf 0.0227} ($2.35\%$) & 	{\bf 0.0189} ($1.96\%$) & 	{\bf 0.0227} ($2.35\%$) & 	{\bf 0.0189} ($1.96\%$) \\ 
  $\log \eta$ & 	 {\bf 0.0174} ($2.04\%$) & 	{\bf 0.0119} ($1.39\%$) & 	{\bf 0.0173} ($2.03\%$) & 	{\bf 0.0118} ($1.39\%$) \\ 
  $\log M_c$ & 	 \bfred{0.0220} ($0.16\%$) & 	\bfred{0.0208} ($0.16\%$) & 	\bfred{0.0220} ($0.16\%$) & 	\bfred{0.0208} ($0.16\%$) \\ 
  $\log \beta$ & 	 0.3397 ($55.53\%$) & 	0.3356 ($54.87\%$) & 	0.3400 ($55.59\%$) & 	0.3359 ($54.91\%$) \\ 
  $A_\mathrm{e,norm}$ & 	 0.2549 ($25.49\%$) & 	0.1365 ($13.65\%$) & 	0.2558 ($25.58\%$) & 	0.1366 ($13.66\%$) \\ 
  $\alpha$ & 	 \bfred{0.0241} ($0.69\%$) & 	\bfred{0.0237} ($0.68\%$) & 	\bfred{0.0241} ($0.69\%$) & 	\bfred{0.0237} ($0.68\%$) \\ 
  \hline
 \end{tabular}
\end{table*}

We \en{now} summarise \en{the} results \en{of} our Fisher analysis on physical parameters, which are relevant to the cross power spectra between haloes and DM.
We examine three hypothetical FRB data with 
$(\bar{n}_\mathrm{FRB}, \Delta \theta_\mathrm{FRB})=(1\, \mathrm{deg}^{-2}, 3')$, $(0.1\, \mathrm{deg}^{-2}, 1')$, and $(10.0\, \mathrm{deg}^{-2}, 10')$ to bracket possible configurations of future FRB surveys.
Note that we set the variance of host-galaxy DM $\sigma_\mathrm{DM} = 60\, \mathrm{pc}/\mathrm{cm}^3$ for every result in this \en{subsection}.

Table~\ref{tab:Fisher_forecast_summary} provides a summary of our Fisher forecasts of six cosmological parameters ($h$, $w_0$, $\Omega_\mathrm{M}$, $\Omega_\mathrm{b}$, $\sigma_8$ and $n_s$), three gas parameters ($\eta$, $M_c$ and $\beta$), \en{one} nuisance parameter for the fraction of free electrons in cosmic baryon density $A_\mathrm{e,norm}$ and \en{one} parameter for the FRB redshift distribution $\alpha$.
In the table, we highlight that the $1\sigma$ constraints within a 1\% and 5\% level \en{of} precision in red and black bold letters, respectively.
Our forecasts are promising.
For a given survey covering $20000 \, {\rm deg}^2$, our Fisher analysis showcases a great potential of the cross power spectra $C_\mathrm{hD}$ as follows: 
\begin{itemize}
\item The cross power spectra can place constraints of $w_0, \Omega_\mathrm{M}$, $\sigma_8$ and $n_s$ with a $\sim5\%$ level \en{provided} 20000 FRBs with an arcmin-level localisation are available. 
\en{Notably, the} expected constraints do not rely on any \en{prior information} from other measurements, e.g. CMB.
\item The cross power spectra enable us to calibrate two astrophysical parameters of $\eta$ and $M_c$ with a level of $1\%$--$2\%$.
The former provides a typical propagation length scale of ejected gas from halo centres, \en{whereas} the latter determines a characteristic halo mass above which more than half of \en{the} gas in the halo is in \en{a} hot and bound state.
\item The parameter constraint with $(\bar{n}_\mathrm{FRB}, \Delta \theta_\mathrm{FRB})=(1\, \mathrm{deg}^{-2}, 3\,\mathrm{arcmin})$
can be similar to the one with $(\bar{n}_\mathrm{FRB}, \Delta \theta_\mathrm{FRB})=(10\, \mathrm{deg}^{-2}, 10\,\mathrm{arcmin})$.
The FRB catalogues with \en{bad} localisation are still able to place meaningful constraints of cosmology and gas physics in a wide range of halo masses, \en{provided} one can set the average number density of FRBs to $\sim10\, \mathrm{deg}^{-2}$.
\item The information of $C_\mathrm{hD}$ at $\ell\sim10^{4}$ is \en{crucial} for precise \en{estimation} of physical parameters 
when the FRB number density is of an order of $\sim0.1\, \mathrm{deg}^{-2}$.
\item The tomographic analysis with halo samples at different redshift bins is efficient to calibrate the redshift distribution of FRBs, although we require some prior knowledge to set a reasonable parametric form of the redshift distribution \citep[see e.g.][]{2021PhRvD.103h3536Q}.
\end{itemize}

Expected constraints of the astrophysical parameters by the cross power spectra $C_\mathrm{hD}$ can provide a powerful test for modelling cosmic baryons with hydrodynamical simulations.
Figure~\ref{fig:mass_limited_gas_params_forecast} shows the expected $1\sigma$ constraints of the three parameters
of $\eta, \beta$ and $M_c$ in our halo model by the cross power spectra.
The figure clearly demonstrates that the limits of the three parameters allow us to distinguish among possible feedback models in the evolution of baryons by cosmological hydrodynamical simulations.

\subsection{Implications}

We \en{now} summarise some implications by our Fisher analysis in \en{Subsection}~\ref{subsec:Fisher_results}.
We study three subjects in modern cosmology and astrophysics:
(i) global mass contents in the late-time universe at $z<1$,
(ii) the gas-to-halo mass relation and (iii) the impact of baryons on total matter density distributions.
To do so, we generate $10^6$ realisations of model parameters following
a multivariate random Gaussian distribution with the covariance being the inverse of the Fisher matrix.
Our model parameters and their central values are listed in Table~\ref{tab:Fisher_summary}.
Using the chain of the random model parameters, 
we evaluate an expected confidence level of some derived parameters,
which are more relevant to subjects (i)--(iii).
When generating the random model parameters, 
we set the Fisher matrix by assuming a hypothetical FRB survey with 
$\bar{n}_\mathrm{FRB}=1\, \mathrm{deg}^{-2}$, $\sigma_\mathrm{DM}=60\, \mathrm{pc}/\mathrm{cm}^3$ and $\Delta \theta_\mathrm{FRB}=3\, \mathrm{arcmin}$.
We include a non-informative prior term in the Fisher matrix (Table~\ref{tab:Fisher_summary}).
We use the information \en{on} a set of the cross power spectra with $\ell_\mathrm{max}=3000$.
See \en{Subsection}~\ref{subsec:foreground_haloes} for the foreground haloes in our hypothetical analysis.

\subsubsection{Global mass contents}



\begin{figure}
 \includegraphics[width=\columnwidth]{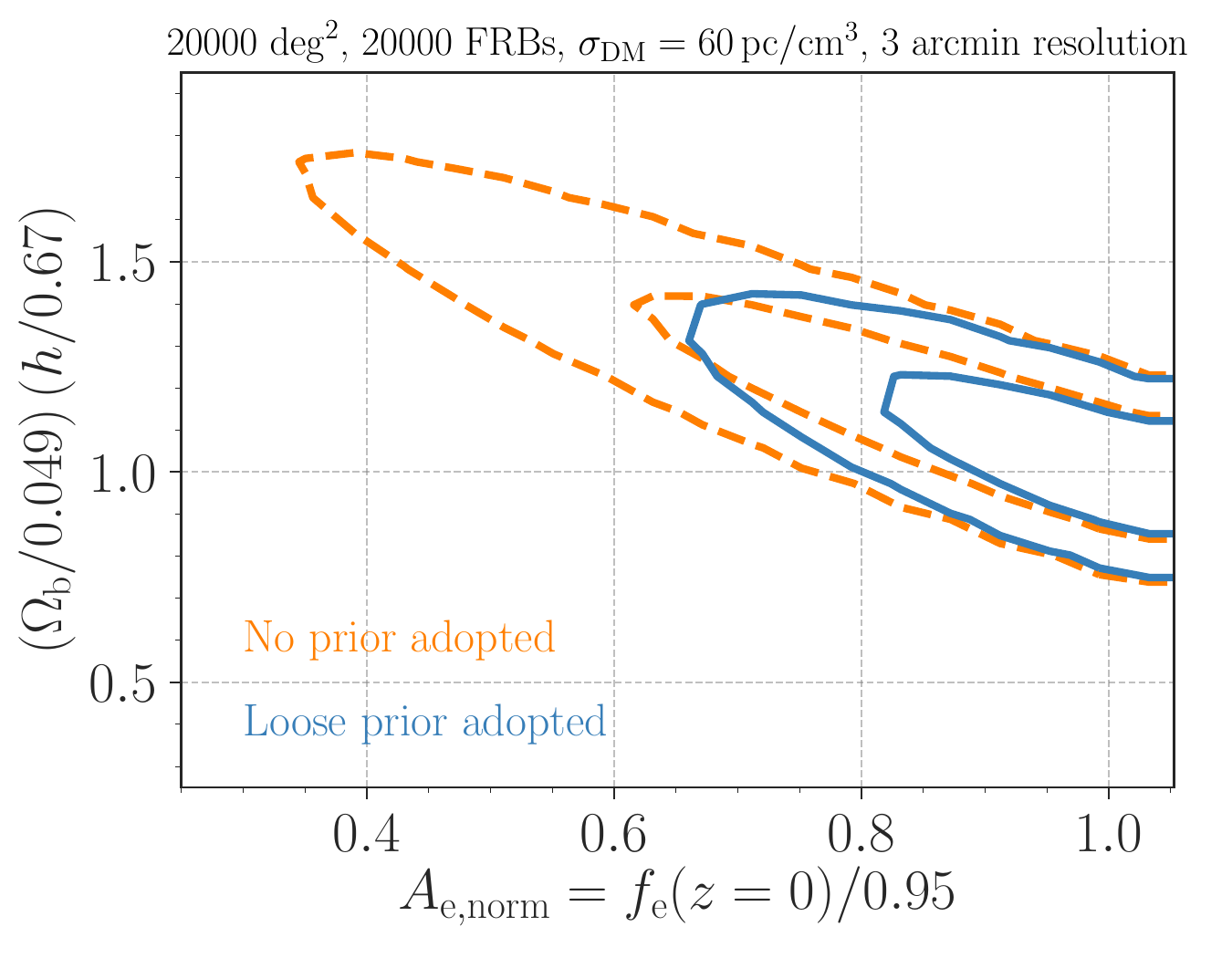}
 \caption{Expected constraint of $\Omega_\mathrm{b}h$ and $A_\mathrm{e,norm}=f_\mathrm{e}(z=0)/0.95$.
 The orange lines show 68\% and 95\% confidence levels derived by our Fisher analysis without any prior information, while the blue lines represent the counterparts with non-informative prior 
 (Table~\ref{tab:Fisher_summary}). Note that the two parameters cannot be simultaneously constrained with the relation of observed DM and host-galaxy redshifts alone \citep[e.g.][]{2020Natur.581..391M}.
 We assume that $20000$ FRBs are available over $20000 \, {\rm deg}^2$ and the variance of host-galaxy DM is set to 60 $\mathrm{pc}/\mathrm{cm}^3$. We also assume that
 the uncertainty of angular positions of individual FRBs is given by 3 arcmin. 
 }
 \label{fig:cosmology_constraint}
\end{figure}

The CMB has been widely used to constrain the average mass density of several components so far; the CMB physics is mainly determined by density fluctuations in the early universe at $z\sim1100$.
Any other measurements of the density fluctuations at lower redshifts 
would provide a powerful test of the standard $\Lambda$ Cold Dark Matter ($\Lambda$CDM) model, allowing \en{us} to explore possible deviations of the $\Lambda$CDM cosmology by new physics \citep[see e.g.][for a review]{2013PhR...530...87W}.
The fluctuations in cosmic mass density at low redshifts can be measured with weak gravitational lensing effects in galaxy images \citep[see e.g.][for a review]{2015RPPh...78h6901K}. 
A primary cosmological parameter in modern weak lensing surveys is given by $S_8 = \sigma_8(\Omega_\mathrm{M}/0.3)^{0.5}$ \citep[e.g.][]{1997ApJ...484..560J, 2021MNRAS.505.4935H}, 
and the latest weak lensing measurements have constrained
$S_8$ with a level of $0.02$--$0.03$ \citep[e.g.][]{2019PASJ...71...43H, 2021A&A...645A.104A, 2021arXiv210513544S}.
As the precision of the lensing measurement increases, a tension \en{appears} between the $S_8$ inferred from the late-time lensing measurements and the counterpart predicted by the CMB constraints.
The origin of this $S_8$ tension is still under debate, while it is obviously \en{imperative} to test the tension with various approaches, which are independent of weak lensing analyses
and CMB.

We expect that the cosmological signal of $C_\mathrm{hD}$ at $\ell \simlt 100$ and its redshift evolution offer a powerful means of determining the linear growth of density perturbations.
Our Fisher analysis predicts that the cross-correlation can place a constraint of $S_8 = 0.851\pm0.012$ at a 68\% confidence level.
This constraint is sufficiently precise to test the $S_8$ tension in modern cosmology.
Note that the constraint of $S_8$ is less dependent on the choice of prior in the cross-correlation analysis.
Even if excluding the non-informative prior, our Fisher analysis shows that $S_8 = 0.851^{+0.012}_{-0.013}$ (68\%) can be realised with the cross-correlation analysis.

The mean baryon density has been precisely determined by the CMB and big-bang nucleosynthesis; \en{these} probes rely on the physics in the primordial photon-baryon plasma.
Measurements of $\Omega_\mathrm{b}$ at low redshifts are always \en{crucial} because \en{they} should provide a valuable check for \en{the} understanding of the baryonic universe.
\citet{2020Natur.581..391M} provided a 10\%-level constraint of $\Omega_\mathrm{b}h$
with the relation between observed DM and host-galaxy redshifts.
Note that the limit in \citet{2020Natur.581..391M} degenerates 
with the fraction of free electrons in cosmic baryon density, $f_\mathrm{e}$ in Eq.~(\ref{eq:mean_ne}).
Hence, other statistical observables of DM would provide some benefits to break the degeneracy among
$\Omega_\mathrm{b}$, $h$ and $f_\mathrm{e}$.

As shown in \en{Subsection}~\ref{subsubsec:param_dependence}, we found that precise measurements of the cross-correlation between DM and dark matter haloes allow \en{breaking} the degeneracy in principle.
To demonstrate the constraining power of the cross-correlation, 
we compute confidence levels in a $\Omega_\mathrm{b}h$-$A_\mathrm{e,norm}$ plane \en{via} the Fisher analysis (see Eq.~[\ref{eq:f_e_model}] for the definition of $A_\mathrm{e,norm}$).
Figure~\ref{fig:cosmology_constraint} shows the expected confidence levels in the $\Omega_\mathrm{b}h$-$A_\mathrm{e,norm}$ plane with the cross-correlation analysis.
The figure illustrates that \en{we} can constrain the two parameters simultaneously, even without any prior on $A_\mathrm{e, norm}$.
Assuming the prior of $A_\mathrm{e,norm}$ with a level of 0.2, we also find that the cross-correlation \en{can} place a constraint of 
$(\Omega_\mathrm{b}/0.049)\, (h/0.67)\, A_\mathrm{e,norm} = 0.993^{+0.030}_{-0.036}$ (68\%).

We also comment on the possibility of constraining the Hubble parameter $h$ with the cross-correlation.
Our fisher analysis shows that a 5\%-level constraint of $h$ can be possible if we impose a non-informative prior (see Table~\ref{tab:Fisher_forecast_summary}). 
This expected constraint is not tight enough to test \en{the} so-called Hubble tension in modern cosmology \citep[e.g.][for a review]{2020NatRP...2...10R}; a combined analysis of the cross-correlation with the DM-redshift relation of $\sim$100 localised FRBs would further improve the constraint \citep[e.g.][]{2021arXiv210800581W}.

\subsubsection{Gas-to-halo mass relations}

The gas mass as a function of \en{the} total mass in a halo, referred to as gas-to-halo mass relation, is a \en{paramount} scaling relation in modern astronomy.
Precise estimates of the gas-to-halo mass relation can improve \en{the} understanding of galaxy formation processes \citep[e.g.][]{2012MNRAS.423.2991V, 2013MNRAS.430.1548H, 2016MNRAS.459.1745F, 2016MNRAS.463.4533V, 2019MNRAS.485.3783D, 2021MNRAS.504.5131L}.
The gas-to-halo mass relation in clusters is of great interest for cluster cosmology \citep[see e.g.][for a review]{2013SSRv..177..247G}.
In addition, the gas-to-halo mass relation in galaxy-sized haloes plays 
an essential role in providing a theoretical interpretation of various observables.
The observables include the thermal SZ effect 
\citep[e.g.][]{2015ApJ...808..151G, 2017MNRAS.467.2315V, 2018PhRvD..97h3501H, 2021ApJ...913...88M}, kinematic SZ effect \citep[e.g.][]{2012PhRvL.109d1101H, 2015PhRvL.115s1301H, 2016PhRvD..93h2002S, 2018MNRAS.475.3764S}, X-ray luminosity \citep[e.g.][]{2010ApJ...709...97L, 2015MNRAS.449.3806A, 2018ApJ...857...32B} and some combinations \citep[e.g.][]{2020ApJ...903...26W, 2021PhRvD.103f3513S}.

In our theoretical model, the gas mass for a halo of $M$ is as \en{follows:}
\beq
M_\mathrm{gas}(<r_\mathrm{out};M) 
= \int_0^{r_\mathrm{out}}\! \mathrm{d}r \, 4\pi r^2\, \left(\rho_\mathrm{BG}(r,M)+\rho_\mathrm{EG}(r,M)\right),
\label{eq:M_gas}
\eeq
where $r_\mathrm{out}$ is a boundary radius, $\rho_\mathrm{BG}$ and $\rho_\mathrm{EG}$ represent the bound and ejected components, respectively (see \en{Subsection}~\ref{subsec:gas_model} for details).
The astrophysical parameters in Table~\ref{tab:model_summary} determine the gas mass $M_\mathrm{gas}$.
Hence, we can infer the gas-to-halo mass relation within our halo-model framework from \en{the} precise measurements of the parameters of $\eta, M_c, \beta$ and $M_{1,0}$.

\begin{figure}
 \includegraphics[width=\columnwidth]{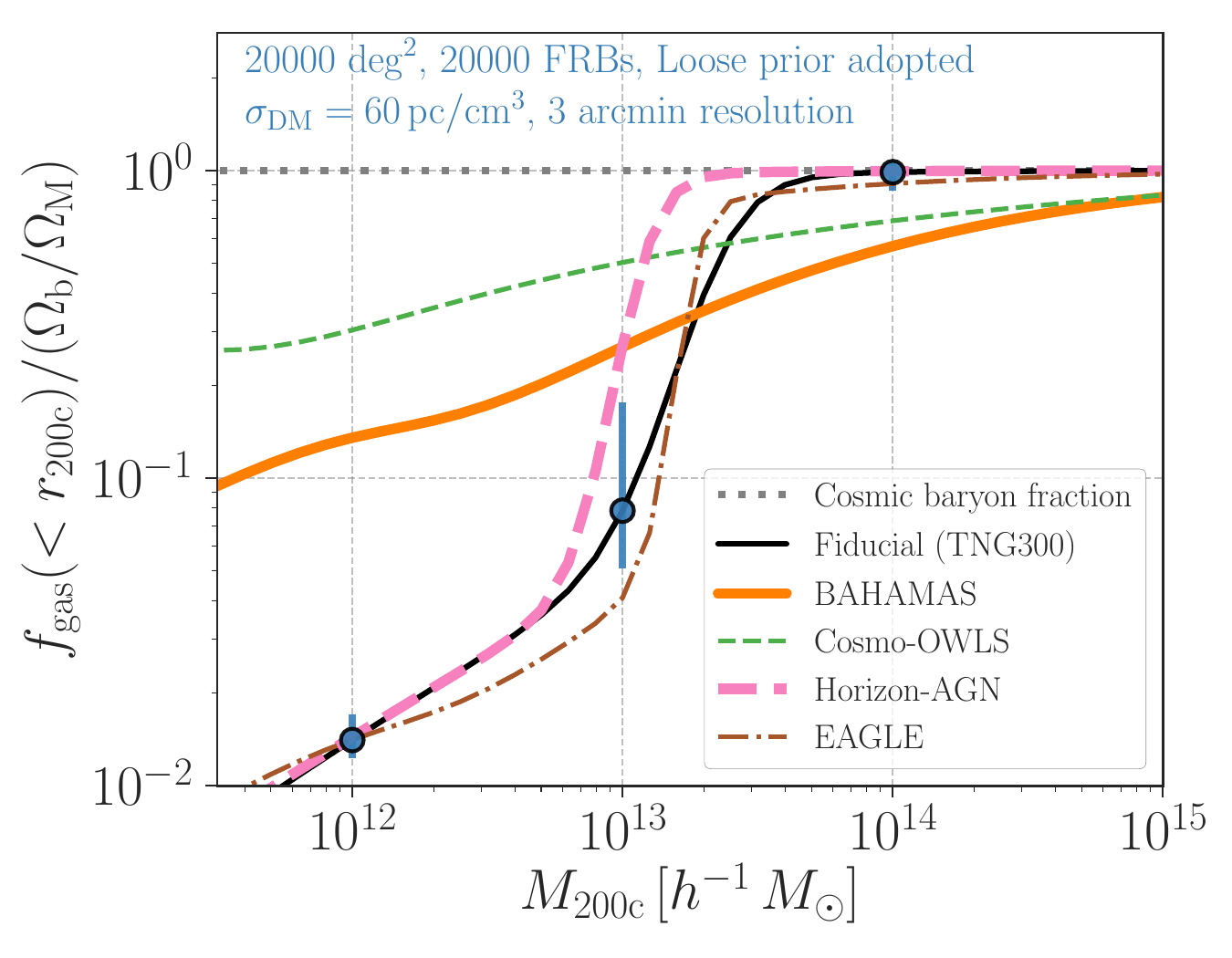}
 \caption{Expected constraints of the gas mass fraction at three halo masses of 
 $M_{200\mathrm{c}}=10^{12}$, $10^{13}$ and $10^{14}\, h^{-1}M_\odot$ at $z=0$.
 We compute the gas mass fraction within the radius of $r_\mathrm{200c}$, divided by the cosmic baryon fraction ($\Omega_\mathrm{b}/\Omega_\mathrm{M}$).
 The blue points with error bars show the limits with a 68\% confidence level, derived via our Fisher analysis of the cross power spectra between dark matter haloes and DM.
 We assume that $20000$ FRBs are available in the sky of $20000 \, {\rm deg}^2$ and set the variance of host-galaxy DM and the localisation error to $60\, \mathrm{pc}/\mathrm{cm}^3$
 and 3 arcmin, respectively. The solid black line shows our fiducial gas-to-halo mass relation expected in the TNG simulation. The other lines represent the relations for the other hydrodynamical simulations.}
 \label{fig:mass_limited_gas_to_halo_mass_forecast}
\end{figure}

Figure~\ref{fig:mass_limited_gas_to_halo_mass_forecast} summarises an expected constraint of the mass fraction of gas in haloes at three halo masses.
In the figure, we define the gas mass fraction as 
$f_\mathrm{gas} = M_\mathrm{gas}(<r_\mathrm{200c})/M_{200\mathrm{c}}$ at $z=0$.
We find that the measurement of $C_\mathrm{hD}$ in our fiducial setup 
can place a constraint of $f_\mathrm{gas} = (2.20^{+0.48}_{-0.30})\times10^{-3}$,
$(1.22^{+1.52}_{-0.43})\times 10^{-2}$ and $0.154^{+0.014}_{-0.020}$
for the haloes with their mass of $M=10^{12}\, h^{-1}M_\odot$,
$10^{13}\, h^{-1}M_\odot$ and $10^{14}\, h^{-1}M_\odot$, respectively.
The above error is set by a 68\% confidence level by our Fisher analysis.
At the mass scale of $10^{13}\, h^{-1}M_\odot$, the error bar in $f_\mathrm{gas}$ becomes larger than two other mass scales because the mass dependence of hot gas fraction ($\beta$ in Table~\ref{tab:model_summary}) cannot be constrained well with our fiducial setup. 
To tighten the error of $\beta$, we would need more bins in halo-mass selections around $M\sim 10^{13}\, h^{-1}M_\odot$.

Lines in Figure~\ref{fig:mass_limited_gas_to_halo_mass_forecast}
show the gas-to-halo mass relations predicted by different hydrodynamical simulations. 
The tight constraint of $f_\mathrm{gas}$ by the measurement of $C_\mathrm{hD}$ enables us to pin down which existing galaxy formation models can explain our baryonic universe.
\en{In addition}, the cross-correlation analysis allows us to calibrate the gas-to-halo mass relation at cluster scales with a level of $\sim 10\%$, providing informative prior for cluster cosmology.
\en{Recalling} that the small-scale cross-correlation can be sensitive to gastrophysics, \en{whereas} the large-scale counterpart is mostly determined by cosmology (see \en{Subsection}~\ref{subsubsec:param_dependence} for details). 
\en{Notably}, the measurement of $C_\mathrm{hD}$ scales with the number density of free electrons, \en{whereas} the X-ray luminosity varies with the density squared and can be subject to the gas clumpiness \citep[e.g.][]{2011ApJ...731L..10N, 2013MNRAS.429..799V, 2015ApJ...806...43B}.

\subsubsection{Sub-grid physics in hydrodynamical simulations}

Although most cosmic mass densities \en{comprise} dark matter, baryons still affect \en{the} statistical properties of total mass density distributions in the universe \citep[see][for a review]{2019OJAp....2E...4C}.
Hydrodynamical simulations are the most accurate \en{techniques for}  predicting the impact of baryons on fluctuations in the cosmic mass density; the simulation itself has a fundamental limitation 
of the resolution of masses or length scales.
To include astrophysical processes associated with scales shorter than the resolution limit, 
the hydrodynamical simulation requires the implementation of various sub-grid recipes.
Some relevant processes include the initial mass function, supernovae and AGN feedback and metal enrichment.
The sub-grid physics has to be calibrated so that the simulation can explain observations, and its impact can be dependent on the assumed cosmology in the simulation.
Because a cosmological hydrodynamical simulation is very time-consuming in general, a flexible method to evaluate the baryonic effects on the matter-density fluctuations is demanded.

Recently, \citet{2020MNRAS.491.2424V} analysed \en{various} hydrodynamical simulations to study how the sub-grid physics recipes affect the matter power spectrum. 
Regardless of details in the recipes, they found a tight correlation (with a $\sim1\%$-level scatter) between 
the mean baryon fraction inside haloes and the power spectrum suppression.
For a comoving wave number of $k$ in units of $h\, \mathrm{Mpc}^{-1}$, the correlation at $z=0$ is approximated as \en{follows:}
\beq
\frac{P_\mathrm{w/\, baryon}}{P_\mathrm{No\, baryon}}-1 
=-\frac{2^a + 2^b \left(c\, \tilde{f}_\mathrm{bar,200c}\right)^{b-a}}{k^{-a}+\left(c\, \tilde{f}_\mathrm{bar,200c}\right)^{b-a}\, k^{-b}}\, \exp\left(d\, \tilde{f}_\mathrm{bar,200c}\, + e\right),
\label{eq:Pm_baryon}
\eeq
where $P_\mathrm{w/\, baryon}$ is the non-linear matter power spectrum with baryonic effects,
$P_\mathrm{No\, baryon}$ is the matter power spectrum expected in dark-matter-only N-body simulations,
$\tilde{f}_\mathrm{bar,200c}$ is 
the baryon mass fraction in haloes of $M_\mathrm{200c} =10^{14}\, M_\odot$ divided by $\Omega_\mathrm{b}/\Omega_\mathrm{M}$,
and fitting parameters are given by
$a=2.111$, $b=0.0038$, $c=1.371$, $d=-5.816$, and $e=-0.4005$.
Note that Eq.~(\ref{eq:Pm_baryon}) holds for the EAGLE, BAHAMAS, Cosmo-OWLS, IllustrisTNG and Horizon-AGN simulations.
\citet{2020MNRAS.495.4800A} also found that the $\tilde{f}_\mathrm{bar,200c}$-dependence in Eq.~(\ref{eq:Pm_baryon}) \en{could} be explained by the gas model in \en{Subsection}~\ref{subsec:gas_model} and the stellar-to-halo mass relation in Appendix~\ref{apdx:f_star}.
\en{Specifically}, the baryon fraction in the model is computed as \en{follows:}
\beq
\tilde{f}_\mathrm{bar,200c} = \left[\frac{M_\mathrm{gas}(<r_\mathrm{200c})}{M} + f_\mathrm{star}(M)\right] \, \left(\frac{\Omega_\mathrm{b}}{\Omega_\mathrm{M}}\right)^{-1}, \label{eq:fb_200c}
\eeq
where $M_\mathrm{200c} = 10^{14}\, M_\odot$, $M_\mathrm{gas}(<r_\mathrm{200c})$ is given by Eq.~(\ref{eq:M_gas}),
and $f_\mathrm{star}(M)$ is the stellar mass fraction provided in Appendix~\ref{apdx:f_star}.

Using the Fisher matrix of our astrophysical parameters ($\eta, M_c, \beta$ and $M_{1,0}$), we found that 
the expected constraint of $\tilde{f}_\mathrm{bar,200c}$ by the measurements of $C_\mathrm{hD}$ is $\tilde{f}_\mathrm{bar,200c} > 0.914\, (0.789)$ with a 68\% (95\%) confidence level.
According to Eq.~(\ref{eq:Pm_baryon}), this lower limit of $\tilde{f}_\mathrm{bar,200c}$ can provide a 
stringent limit of $|P_\mathrm{w/\, baryon}/P_\mathrm{No\, baryon}-1|\simlt 0.03$
at $k=1\, h\, \mathrm{Mpc}^{-1}$ at the 95\% confidence level.
Hence, the cross power spectrum with dark matter haloes and FRBs has a great potential of constraining the baryonic effect on the small-scale matter power spectrum, which is the main theoretical uncertainty in weak lensing cosmology today \citep[see e.g.][for a review]{2018ARA&A..56..393M}.

\section{Limitations}\label{sec:limitations}

Before concluding, we summarise the major limitations 
in our correlation statistics between dark matter haloes and cosmic \en{DMs}. 
The following issues will be addressed in \en{future} studies.

\subsection{Precise calibration of halo model}\label{subsec:model_accuracy}

As shown in \en{Subsection}~\ref{subsec:comp_model_TNG} and Appendix~\ref{apdx:comp_TNG},
our model of gas in a halo can provide a fit to the power spectra in the TNG simulation within a level of $\sim25\%$ over a wide range of angular scales, halo masses and redshifts.
Nevertheless, the $25\%$-level precision in our model is \en{insufficient}  to obtain unbiased estimates of the cosmological and astrophysical parameters from future measurements of $C_\mathrm{hD}$.
We \en{now} list points to be \en{considered} for more precise models of $C_\mathrm{hD}$ below.

Provided that the localisation error of FRBs is set to $O(1)\, \mathrm{arcmin}$,
the smallest relevant comoving scale to $C_\mathrm{hD}$ is 
of an order of $\sim 200 \, \mathrm{kpc}$ at $z=0.1$ and becomes larger as $z$ \en{increases}.
For galaxy-sized haloes, the scale of $200\, \mathrm{kpc}$ provides a typical expansion of the circumgalactic medium \citep[see e.g.][for a review]{2017ARA&A..55..389T}.
On the other hands, $200 \, \mathrm{kpc}$ corresponds to a central region of cluster-sized haloes.
Gas cooling due to some strong feedback at the cluster core must be accounted for, while various observational information about the `warm' circumgalactic medium at the edge of galaxy-sized haloes would be useful \citep[e.g.][]{2018MNRAS.478.2909S, 2018AJ....156...66M}.

When applying actual data of galaxies and clusters, we also require 
more realistic recipes to include satellite galaxies and 
the scatter in mass-to-observable relations for galaxy clusters.
A halo occupation model \citep[e.g.][]{2003ApJ...593....1B} and/or
forward modelling approach \citep[e.g.][]{2005PhRvD..72d3006L}
would be promising.
Further, the scatter in the gas-to-halo mass relation can be linked to 
some secondary halo parameters, such as shapes, ages and mass accretion rates \citep[e.g.][for a cluster scaling relation]{2020MNRAS.496.2743G}.
Our model cannot account for possible correlations between the selection function of haloes and secondary halo parameters.

On larger scales beyond the virial regions of haloes,
the clustering of two separated haloes mainly contributes to the expected signal of $C_\mathrm{hD}$.
We adopt a linear bias to compute the clustering signal as in Eq.~(\ref{eq:2halo_hh}), but it is an \en{inaccurate} approximation at scales of $O(10\, \mathrm{Mpc})$ \citep[e.g.][]{2021MNRAS.503.3095M}.
At sufficient large scales, our model predicts that 
the signal of $C_\mathrm{hD}$ scales with $\sim b_\mathrm{h}\, f_\mathrm{e}$, where $b_\mathrm{h}$ is the linear halo bias and 
$f_\mathrm{e}$ is the fraction of free electrons in the cosmic baryon number density.
The SNR of the clustering signal can reach $\sim10$--$50$ if 20000 FRBs in a sky coverage of $20000 \, \mathrm{deg}^2$ are available (see Figure~\ref{fig:mass_limited_s2n_eachz_eachm}).
This indicates that we would require a model of $f_\mathrm{e}$ with a $2\%$-level precision, but our model of $f_\mathrm{e}$ still relies on phenomenological fitting functions based on the TNG simulation \citep{2021MNRAS.502.2615T}.
A more physically motivated description of $f_\mathrm{e}$ is demanded to \en{consider} some galactic feedback processes.
\en{Notably}, our model cannot fully account for warm-hot intergalactic medium (WHIM) in cosmic filaments \citep{1999ApJ...514....1C}.
The ejected gas components in our model may be responsible for some parts of WHIM; we completely ignore the accretion of flesh gas through the filaments.

\subsection{Realistic statistical errors}

We employ a Gaussian approximation to compute the statistical errors of 
the cross power spectrum $C_\mathrm{hD}$.
Non-linear gravitational growth in cosmic mass density can induce a complex coupling 
of density fluctuations with different Fourier modes.
The mode coupling effect causes a four-point correlation among relevant observables
and increases the statistical errors \citep[e.g.][]{2013PhRvD..87l3504T}.
For example, \citet{2021arXiv210713692C} claimed the detection of a large average DM from nearby dark matter haloes of galaxies at distances of $<40\, \mathrm{Mpc}$.
Such nearby objects can induce additional scatters of our cross-correlation and affect the parameter inference in principle.
The additional sample variance can be interpreted as the four-point correlation arising from Poisson number fluctuations of galaxies in the survey window \citep[see e.g.][for other cosmological observables]{2009ApJ...701..945S, 2021PhRvD.103f3501O}.
Although the four-point correlation can be computed in a halo-model approach, we leave its impact on parameter forecasts for future studies.

In practice, the localisation error of FRBs may depend on the sky coordinate of FRBs, making the function of Eq.~(\ref{eq:beam}) more complicated \citep[e.g.][]{2021ApJ...922...42R}.
More realistic treatments of the localisation error would be needed in actual measurements of $C_\mathrm{hD}$.
The redshift distribution of FRBs is a key ingredient in the cross-correlation analysis.
A set of localised FRBs with sub-arcseconds is needed to 
have a reasonable estimate of the redshift distribution.
Target selection in a small sky coverage may be subject to the sample variance \citep[see e.g.][for similar studies in the light of photometric redshifts in galaxy imaging surveys]{2012MNRAS.423..909C}.
It would be interesting to study how many FRBs have to be localised with a level of sub-arcseconds and how wide sky coverage is required for a precise inference of the redshift distribution.
The one-point probability distribution of observed \en{DMs} can provide complementary information about the redshift distribution of FRBs \citep{2017PhRvD..95h3012S}.
\en{In addition}, we may use another FRB observable of the scattering time\footnote{A FRB pulse can be broadened because of multi-path propagation by scattering \citep[e.g.][]{1985MNRAS.213..591B}.
The scattering time provides a typical time scale of the temporal broadening of FRBs by the scattering due to electron density variations along a line of sight.} to control the variance of the host-galaxy DM \citep[e.g.][]{2021arXiv210710858C, 2021arXiv210801172C}, 
but more investigations would be required.

\subsection{DMs from FRB sources}\label{subsec:D_source}

We \en{have ignored} possible contributions from interstellar media or stellar discs at FRB sources 
(i.e., basically from the host galaxies) to the cross-correlation analysis, although this contribution is largely dependent on the origin of FRBs. 
We \en{now} provide a rough order-of-magnitude estimate based on a halo model.

We first assume that \en{the} physical sizes of FRB sources are small enough to be approximated as point sources.
This assumption is expected to be valid \en{provided} FRBs are associated with astrophysical compact objects with a short duration of $\sim1\, \mathrm{ms}$ \citep[e.g.][]{2020ApJ...895L..37B}, 
and we work with scales larger than $\sim 100\, \mathrm{kpc}$.
Then, we write cumulative DMs from FRB sources as \en{follows:}
\beqa
D_\mathrm{source}(\bmtheta) \!\!\!\!&=&\!\!\!\!
\frac{1}{\cal{N}_\mathrm{FRB}} \int_0^{\chi_H}\! \mathrm{d}\chi\, \chi^2\, f_\mathrm{duty}(\chi)\, \int \! \mathrm{d}M\, \frac{\mathrm{d}n}{\mathrm{d}M}(M,\chi)\, 
\nonumber \\
&&
\qquad
\times 
\, P\left(M,z(\chi)\right) \bar{\tau}_\mathrm{s}\, \left[1+\delta_\mathrm{h}(\chi\bmtheta, \chi; M)\right], \label{eq:Dsource} \\
\cal{N}_\mathrm{FRB} \!\!\!\!&=&\!\!\!\! \int_0^{\chi_H}\! \mathrm{d}\chi\, \chi^2\, f_\mathrm{duty}(\chi)\, \int \! \mathrm{d}M\, \frac{\mathrm{d}n}{\mathrm{d}M}(M,\chi)\, P\left(M,z(\chi)\right),
\eeqa
where $P(M,z)$ represents the number of FRB sources in a halo of $M$ at the redshift $z$, $\bar{\tau}_\mathrm{s}$ is the column density of free electrons around single FRBs
and $f_\mathrm{duty}(\chi)$ 
provides the fraction of {\it observed} FRBs to their total number.
For simplicity, we ignore any dependence of $\bar{\tau}_\mathrm{s}$ on halo masses, redshifts and other properties.
\en{Notably}, $\bar{\tau}_\mathrm{s}$ should be evaluated by the average column density over \en{the} orientation of stellar discs in FRB host haloes.

In Eq.~(\ref{eq:Dsource}), we can normalise $P(M,z)$ by $\int \mathrm{d}M\, P(M,z)=1$ at each redshift, without loss of generality.
Using recent observational constraints in \citet{2020ApJ...903..152H}, 
we find that $P(M,z)$ can be approximated as a log-normal distribution in terms of stellar masses\footnote{Note that $P(M,z)$ is poorly understood yet. A recent observation implies that a typical host halo mass of FRBs may be cluster-sized \citep{2021ApJ...922...42R}, but more observations are required to make a robust conclusion.}:
\beqa
P(M,z) \!\!\!\!&=&\!\!\!\! P(M_\mathrm{star}(M,z)) \, \frac{\mathrm{d}M_\mathrm{star}}{\mathrm{d}M}, \label{eq:P_M_source}\\
P(M_\mathrm{star}) \!\!\!\!&=&\!\!\!\! \frac{1}{M_\mathrm{star}\ln 10}\frac{1}{\sigma\sqrt{2\pi}}
       \exp\left[-\left(\frac{\log M_\mathrm{star}-\mu}{2\sigma}\right)^2\right],
\eeqa
where $M_\mathrm{star}(M,z)$ is the stellar mass (in units of $M_\odot$) in a halo $M$ at $z$, $\mu=9.30$ and $\sigma=0.77$. 
The stellar-to-halo mass relation is given in Appendix~\ref{apdx:f_star}.
Then, the fraction of $f_\mathrm{duty}$ is determined so that the redshift distribution of FRBs follows Eq.~(\ref{eq:pz}).
\en{Specifically}, we set $f_\mathrm{duty}$ by
\beq
p(\chi) = \chi^2\, f_\mathrm{duty}(\chi)\, \int \!\mathrm{d}M\, \frac{\mathrm{d}n}{\mathrm{d}M}\, P\left(M,z(\chi)\right),
\eeq
where the normalisation is given by $\int \mathrm{d}\chi\, p(\chi) = 1$.

The halo model predicts that the cross-correlation between $D_\mathrm{source}$ and $\delta^\mathrm{2D}_\mathrm{h}$ 
scales with $\bar{\tau}_\mathrm{s}$ and 
\beqa
f_\mathrm{eff, FRB} \!\!\!\!&=&\!\!\!\!
\frac{1}{\cal{N}_\mathrm{FRB}}\int_{\chi_{l,\mathrm{min}}}^{\chi_{l,\mathrm{max}}}\! \mathrm{d}\chi\, \chi^2\, f_\mathrm{duty}(\chi) \nonumber \\
\!\!\!\!&&\!\!\!\!
\quad
\times \int\! \mathrm{d}M \frac{\mathrm{d}n}{\mathrm{d}M}(M,\chi)\, S(M)\, P(M,z(\chi)), \label{eq:feff_FRB}
\eeqa
where $f_\mathrm{eff, FRB}$ represents the fraction of FRB host haloes \en{that} share the selection of halo masses and redshifts with $\delta^\mathrm{2D}_\mathrm{h}$.
For mass-limited samples with $M\ge 10^{12}\, h^{-1}M_\odot$,
our fiducial model with Eq.~(\ref{eq:P_M_source}) predicts 
that $f_\mathrm{eff, FRB}/10^{-4} = 0.37, 1.36, 1.81, 1.69$ and $1.32$ at $0<z\le 0.2$, $0.2<z\le 0.4$, $0.4<z\le 0.6$, $0.6<z\le 0.8$ and $0.8<z\le 1.0$, respectively.
On the other hand, we find that $f_\mathrm{eff, FRB}$ is smaller than $10^{-7}$ for mass-limited samples with $M\ge 10^{14}\, h^{-1}M_\odot$ for all relevant redshift bins.

For comparison, the \en{DM} by free electrons distributed in a halo\footnote{We evaluate this as
$(X_\mathrm{p}+Y_\mathrm{p}/2)\, \left[f_\mathrm{gas}\, M\, (2 r_\mathrm{200c})\right]\, \left[(4/3)\, \pi \, r^3_\mathrm{200c}\, m_\mathrm{p}\right]^{-1}$.} of $M$ 
in a narrow redshift range can roughly scale with
\beq
420\, \mathrm{pc}/\mathrm{cm}^3 \left(\frac{f_\mathrm{gas}(M,z)}{0.15}\right)\, \left(\frac{M}{10^{14}\, M_\odot}\right)^{1/3}\,
\left(\frac{H(z)}{100\, \mathrm{km}/\mathrm{s}/\mathrm{Mpc}}\right)^{4/3}, \label{eq:Dex_eachhalo}
\eeq
where $f_\mathrm{gas}$ is the mass fraction of gas in a halo,
and it depends on the halo mass and redshift.

Suppose that a typical value of $\bar{\tau}_\mathrm{s}$ ranges from $100$--$1000\, \mathrm{pc}/\mathrm{cm}^3$, we expect $f_\mathrm{eff, FRB} \bar{\tau}_\mathrm{s} \ll 1\, \mathrm{pc}/\mathrm{cm}^3$ for cluster-sized haloes, while $f_\mathrm{eff, FRB} \bar{\tau}_\mathrm{s} \sim 0.01-0.1\, \mathrm{pc}/\mathrm{cm}^3$ can be possible for galaxy-sized haloes.
Hence, the cross-correlation for cluster-sized haloes 
is less affected by the near-source plasma, 
allowing us to perform a robust cosmological analysis.
To constrain astrophysical parameters, we require \en{a} cross-correlation function with less massive haloes.
Because Eq.~(\ref{eq:Dex_eachhalo}) gives $\sim 0.9\, \mathrm{pc}/\mathrm{cm}^{3}$ 
at the mass of $\sim10^{12}\, M_\odot$ in our fiducial scenario 
($f_\mathrm{gas}/0.15 \simeq 10^{-2}$ at $M\sim10^{12}\, M_\odot$), 
the near-source plasma with $f_\mathrm{eff, FRB} \bar{\tau}_\mathrm{s} \sim 0.01-0.1\, \mathrm{pc}/\mathrm{cm}^3$ may affect the cross-correlation for galaxy-sized haloes.
Thus, we expect that the near-source plasma can be a major systematic source to constrain gastrophysics. 
Future studies need to develop a more realistic model of $D_\mathrm{source}$ by incorporating \en{the} physics of compact objects.

\subsection{Possible extensions}\label{subsec:possible_ext}

\rev{So far we have worked with the single projected DM field, but this choice is not optimal to extract the information in the halo-DM cross correlations.}
\rev{One can use the observed DM values of individual FRBs as an indicator of their redshifts and divide the FRBs into several bins with their DM estimates \citep[e.g.][]{2020PhRvD.102b3528R, 2021ApJ...922...42R}.
Such a DM-binned analysis would be helpful not only to constrain the underlying distribution of FRB redshifts, but also to mitigate possible systematic effects due to correlations between haloes and FRB sources in same redshift bins (see Subsection~\ref{subsec:D_source}).}

\rev{When dividing FRBs with their DMs, we have to include the contribution of DMs within FRB host galaxies (i.e. $D_\mathrm{source}$) in our model properly. At this early stage, the modelling of $D_\mathrm{source}$ cannot be as precise as our halo model of large-scale structures. Another issue in the analysis with DM bins is a modulation effect of the observed FRB number density as pointed out in \citet{2020PhRvD.102b3528R}. Because free electrons in the large-scale structure are inhomogeneous, the redshift distribution of FRB sources should depend on angular coordinates after one imposes the selection of observed DMs. This modulation effect makes our halo-model formulation more complicated.
Furthermore, potential selection biases in FRB observations (e.g.~selection against high-DM FRBs) can severely affect the cross-correlation analysis with the binning of DMs. Hence, we restrict ourselves to the single projected DM field in the present paper, while it would be interesting to explore how much additional information can be extracted with DM-binned cross correlation functions.}

\rev{We also assume that precise redshift catalogues of galaxies and clusters are available in the analysis, but it would be important to include realistic redshift errors of foreground halo catalogues in the halo model for future applications, especially when one would work 
on galaxy imaging data with photometric redshifts.
Interesting candidates of galaxy imaging surveys include
the ten-year Rubin Observatory Legacy Survey of Space and Time (LSST)\footnote{\url{https://www.lsst.org/}} and \textit{Euclid}\footnote{\url{https://www.euclid-ec.org/}}.
In general, photometric redshift sample would have a denser number density than the spectroscopic counterpart, allowing to reduce shot-noise errors in small-scale cross correlations.}

\section{Conclusions and discussions}\label{sec:conclusions}

In this \en{article}, we studied information contents in cross-correlation analyses with dark matter haloes and \en{DMs} arising from large-scale structures.
Near-future observations of FRBs will allow us to perform a statistical analysis of the DM over many lines of sight.
The cross-correlation of the DM and the position of dark matter haloes 
provides a powerful means of studying the redshift evolution in free electrons \en{and} the non-linear baryonic process of gas in single dark matter haloes.
We adopted a halo-model approach to predict an expected cross-correlation signal of the DM and several mass-limited halo samples.
Our halo model assumes a two-phase gas scenario, which \en{had} been calibrated with a set of hydrodynamical simulations \citep{2020MNRAS.495.4800A}.
We also \en{considered} a realistic redshift evolution of the average number density in free electrons, which is consistent with the IllustrisTNG simulations \citep{2018MNRAS.475..676S, 2021MNRAS.502.2615T}.
Our model has six parameters to set \en{the} cosmological evolution of dark matter haloes, and four parameters to characterise the gas density profile in single haloes, the gas-to-halo mass relation and the stellar-to-halo mass relation.

According to a rich phenomenology in the halo model, we improved our previous results \citep{2017PhRvD..95h3012S} by accounting for non-linear gravitational and baryonic effects in the cross-correlation analysis.
Our findings in this \en{study} are summarised below.

\begin{enumerate}

\item Our halo model predicts that \en{a} cross power spectrum with cluster-sized haloes at $\ell \simlt 10^4$ is sensitive to the cosmological parameters, but it is less affected by \en{gastrophysics}. 
\en{Moreover, a} cross power spectrum with group- and galaxy-sized haloes would contain the information \en{about} baryonic physics at angular scales of $\simlt 10\, \mathrm{arcmin}$. Therefore, a combined analysis of the cross-correlation with haloes at different halo masses would be efficient to constrain cosmology and astrophysics separately.

\vspace{12pt}

\item The normalisation of the cross power spectrum scales with a combination of $\Omega_\mathrm{b}\, h\, f_\mathrm{e}$ ($\Omega_\mathrm{b}$, $h$ and $f_\mathrm{e}$ are
the mean baryon density, the present-day Hubble parameter and the fraction of free electrons in cosmic baryon density today, respectively), \en{whereas} the shape of the power spectrum is affected by $\Omega_\mathrm{b}$ and $h$ but not by $f_\mathrm{e}$. Hence, a detailed analysis of the cross-correlation can break the primary degeneracy of $\Omega_\mathrm{b}\, h\, f_\mathrm{e}$ in principle.

\vspace{12pt}

\item Assuming that 20000 FRBs are available in a sky coverage of $20000\, \mathrm{deg}^2$ with a plausible scatter of the DM in FRB host galaxies and the localisation error of FRBs being $3\, \mathrm{arcmin}$, we found that the SNR of the cross power spectrum at $\ell \simlt 3000$ 
can reach $\sim100$ for haloes in a wide range of masses and redshifts. 
This indicates that the future FRB data \en{will} allow us to measure the cross-correlation signal with a level of $\sim 1/100\sim 1\%$ precision.

\vspace{12pt}

\item We performed a Fisher analysis to forecast parameter constraints with cross-correlation analyses with dark matter haloes and the DM from the future FRB data. We found that the cosmological parameters \en{could} be constrained with a level of 5\%, even if informative priors from the \en{CMB are excluded}.
On the astrophysical parameters, the cross-correlation analysis \en{could} put a tight constraint of a typical propagation length scale of ejected gas from halo centres and 
a characteristic halo mass below which more than half of the gas in the halo comprises the ejected component.

\vspace{12pt}

\item The expected parameter constraints by the cross-correlation can shed light \en{on} the tension of gravitational growth of cosmic matter density in modern cosmology because the cross-correlation with different redshift bins is efficient to trace the redshift evolution of matter density fluctuations. 
We expect that the cross-correlation has a potential of constraining $S_8 = \sigma_8 (\Omega_\mathrm{M}/0.3)^{0.5}$ with a level of $S_8 = 0.851\pm 0.012$ (68\% CL), where $\Omega_\mathrm{M}$ and $\sigma_8$ represent the mean cosmic mass density and the linear mass variance smoothed by $8\, h^{-1}\mathrm{Mpc}$, respectively.
In addition, the amplitude of the cross-correlation function can place a meaningful limit of the mean baryon density at low redshifts. The expected limit by the future cross-correlation is
$(\Omega_\mathrm{b}/0.049)\, (h/0.67)\, (f_\mathrm{e}/0.95) = 0.993^{+0.030}_{-0.036}$ at a 68\% confidence level.
These limits would provide an important consistency test of structure formation in an expanding universe based on low-redshift information alone.

\vspace{12pt}

\item The cross-correlation with haloes and DM in future FRB surveys can also place a stringent limit \en{on} the gas-to-halo mass relation and the baryon content (i.e. the sum of stars and gas) in cluster-sized haloes.
The expected limit on the gas-to-halo mass relation allows us to 
pin down which existing galaxy formation models can explain our baryonic universe.
The stringent limit of the baryon content in clusters would provide a reasonable estimate on the strength of feedback by AGN, bracketing baryonic effects on \en{spatial} clustering in large-scale structures.
\end{enumerate}

Our results provide a major step forward \en{to} precision cosmology with FRBs, \en{but} there are some limitations to apply our method to future datasets.
We require a more precise model for the cross-correlation between haloes and DM.
Our halo model can reproduce the cross power spectra for mass-limited haloes in the TNG simulations with a level of $25\%$. 
Note that the gas model in this study has been calibrated so that it can provide a reasonable fit to non-linear matter power spectra in simulations.
We expect that a direct calibration of gas density profiles around single 
haloes in the simulations can improve the model accuracy but leave it for future studies.
In addition, more observational studies \en{are crucial} in determining the redshift distribution of FRBs at a given observational condition. 
We adopt a simple parametric model of the redshift distribution, but this has to be improved before a large sample of FRBs would be available.
Further, we ignore possible contributions from near-source plasma in \en{the} interstellar medium 
(i.e., basically from the host galaxy) to the cross-correlation analysis. 
We provided a rough order-of-magnitude estimate of the near-source plasma in \en{Subsection}~\ref{subsec:D_source}, but more realistic modelling associated with astrophysics of compact objects will be needed for future studies.

\section*{acknowledgements} 
This work is supported by MEXT KAKENHI Grant Number (19K14767, 20H04723, 20H05855, 20H05861)
and Grant-in-Aid for JSPS Fellows Grant Number JP21J00011.
This work is also supported by MEXT KAKENHI No. 20H01901, 20H01904, 20H00158, 18H01213, 18H01215, 17H06357, 17H06362, and 17H06131 (KI). 
We also thank the participants of the workshops with the identification number YITP-T-20-04 for discussions.
KO is supported by JSPS Research Fellowships for Young Scientists.
Numerical computations were in part carried out on Cray XC50 at Center for Computational Astrophysics, National Astronomical Observatory of Japan.

\section*{Data Availability}
The data underlying this article will be shared on reasonable request to the corresponding author.

\begin{table*}
 \caption{
 Variables used in this paper.
 }
 \label{tab:variables}
 \begin{center}
 \begin{tabular}{@{}lll}
  \hline
  Symbols & Definition & Eqs.\\
  \hline
  $H(z)$ & Hubble parameter at redshift $z$ & (\ref{eq:Hz})\\
  $\chi(z)$ & Radial comoving distance to $z$ & (\ref{eq:chi_z})\\
  $\bar{n}_\mathrm{h}^\mathrm{2D}$ & Average surface number density of haloes & (\ref{eq:nhalo_2D}) \\
  $\delta_\mathrm{h}^\mathrm{2D}$ & Projected number density fluctuation of haloes & (\ref{eq:nhalo_2D}) \\
  $\mathrm{d}n/\mathrm{d}M$ & Halo mass function & (\ref{eq:delta_h_2D}) \\
  $S(M, \chi)$ & Selection function of haloes with $M$ at $\chi$ to define $\delta_\mathrm{h}^\mathrm{2D}$ & (\ref{eq:delta_h_2D}) \\
  $D_\mathrm{MW}$ & Dispersion measure (DM) in the Milky Way & (\ref{eq:obs_DM}) \\
  $D_\mathrm{LSS}$ & DM from electrons in large-scale structures & (\ref{eq:obs_DM}), (\ref{eq:D_LSS_singlez}) \\
  $D_\mathrm{source}$ & Host-galaxy DM & (\ref{eq:obs_DM}), (\ref{eq:Dsource})\\
  $f_\mathrm{e}$ & Fraction of free electrons in the cosmic electron number density & (\ref{eq:mean_ne}) \\
  $p_z(z)$ & Redshift distribution of FRBs & (\ref{eq:pz}) \\
  $\xi_\mathrm{hD}$ & Two-point cross-correlation between observed DMs and $\delta_\mathrm{h}^\mathrm{2D}$ & (\ref{eq:cross_2pcf}) \\
  $C_\mathrm{hD}$ & Cross power spectrum between observed DMs and $\delta_\mathrm{h}^\mathrm{2D}$ & (\ref{eq:def_Cl}) \\
  $\delta_\mathrm{D}^{(n)}$ & $n$-dimensional Dirac delta function & (\ref{eq:def_Cl}) \\
  $P_\mathrm{he}$ & Power spectrum between free electron and halo density fields in 3D & (\ref{eq:P_he}) \\
  $n_\mathrm{e,h}$ & A spherical electron density profile around a halo & (\ref{eq:P1h})-(\ref{eq:hm_norm}) \\
  $\rho_\mathrm{BG}$ & Mass density profile of a hot gas component & (\ref{eq:rho_BG}) \\
  $f_\mathrm{BG}$ & Mass fraction of the hot gas component & (\ref{eq:frac_BG}) \\
  $f_\mathrm{star}$ & Stellar mass fraction & (\ref{eq:frac_BG}), (\ref{eq:frac_star}) \\
  $\rho_\mathrm{EG}$ & Mass density profile of a gas ejected from a halo & (\ref{eq:rho_EG}) \\
  $f_\mathrm{EG}$ & Mass fraction of the ejected gas & (\ref{eq:frac_EG}) \\
  $\Delta \theta_\mathrm{FRB}$ & Angular resolution of FRB postion's & (\ref{eq:beam}) \\
  ${\cal B}$ & Smearing effect on $C_\mathrm{hD}$ due to position errors of FRBs & (\ref{eq:beam}) \\
  $C_\mathrm{DD}$ & Power spectrum of observed DM & (\ref{eq:c_dd}) \\
  $\sigma_\mathrm{DM}$ & Variance of host-galaxy DM & (\ref{eq:c_dd}) \\
  $\bar{n}_\mathrm{FRB}$ & Average surface number density of FRBs & (\ref{eq:c_dd}) \\
  $C_\mathrm{hh}$ & Power spectrum of $\delta_\mathrm{h}^\mathrm{2D}$ & (\ref{eq:c_hh}) \\
  $F_{ij}$ & Fisher matrix & (\ref{eq:Fisher}) \\
  $\mathrm{S}/\mathrm{N}$ & Signal-to-noise ratio & (\ref{eq:s2n}) \\
  $M_\mathrm{gas}$ & Gas mass of a halo & (\ref{eq:M_gas}) \\
  $P_\mathrm{w/\, baryon}$ & Non-linear matter power spectrum with baryonic effects & (\ref{eq:Pm_baryon}) \\
  $P_\mathrm{No\, baryon}$ & Non-linear matter power spectrum without baryonc effects & (\ref{eq:Pm_baryon}) \\
  $\tilde{f}_\mathrm{bar, 200c}$ & Baryon mass fraction in haloes with their mass of $10^{14}\, M_\odot$  & (\ref{eq:Pm_baryon}), (\ref{eq:fb_200c}) \\
  $P(M,z)$ & Selection function of FRBs as a function of halo mass $M$ at redshift $z$ & (\ref{eq:Dsource}), (\ref{eq:P_M_source}) \\
  $f_\mathrm{eff, FRB}$ & Fraction of FRB hosts that share the mass/redshift selection with $\delta_\mathrm{h}^\mathrm{2D}$ & (\ref{eq:feff_FRB}) \\
  \hline
 \end{tabular}
 \end{center}
\end{table*}


\appendix

\section{A model of stellar-to-halo mass relation}\label{apdx:f_star}

In this appendix, we summarise a model of stellar masses as a function of 
halo masses and redshifts. We adopt the model in \citet{2020MNRAS.495.4800A}, which 
follows the parametrisation from the abundance matching analysis in \citet{2013ApJ...770...57B}.

The stellar mass fraction in a halo is given by
\beq
f_\mathrm{star}(M, z) = \epsilon\, \left(\frac{M_1}{M}\right)\,
10^{g(\log M_1/M)-g(0)}, \label{eq:frac_star}
\eeq
where 
\beq
g(x) = -\log(10^{\alpha x}) + \delta\, \frac{\left[\log\left(1+e^{x}\right)\right]^{\gamma}}{1+\exp\left(10^{-x}\right)},
\eeq
and $\alpha$, $\delta$, $\gamma$, $\epsilon$ and $M_1$ are 
redshift-dependent parameters in the model.
We set the model parameters at $z=0$ with the results in \citet{2018AstL...44....8K},
while the redshift dependence is assumed to be the same as in \citet{2013ApJ...770...57B}.
To be specific, the parameters are given by
\beqa
\nu(a) \!\!\!\!&=&\!\!\!\! \exp(-4a^2),\\
\log M_1 \!\!\!\!&=&\!\!\!\! \log(M_{1,0}) + m_{1,a}(a-1) + m_{1,z}z, \\
\log \epsilon \!\!\!\!&=&\!\!\!\! \log\left(\epsilon_0\right) + \epsilon_a (a-1)\nu + \epsilon_{a,2}(a-1), \\
\alpha \!\!\!\!&=&\!\!\!\! \alpha_0 + \alpha_a (a-1)\nu, \\
\delta \!\!\!\!&=&\!\!\!\! \delta_0 + \left[\delta_a \left(a-1\right) + \delta_z z\right]\nu, \\
\gamma \!\!\!\!&=&\!\!\!\! \gamma_0 + \left[\gamma_a \left(a-1\right) + \gamma_z z\right]\nu,
\eeqa
where $a=1/(1+z)$ and 
$m_{1,a}=-1.793$, $m_{1,z}=-0.251$, 
$\epsilon_0 = 0.023$, $\epsilon_a = -0.006$, $\epsilon_{a,2}=-0.119$, 
$\alpha_0 = -1.779$, $\alpha_a = 0.731$,
$\delta_0 = 4.394$, $\delta_a = 2.608$, $\delta_z = -0.043$,
$\gamma_0 = 0.547$, $\gamma_a = 1.319$, and $\gamma_z = 0.279$.

In this paper, we regard $M_{1,0}$ as a single free parameter in the model and fix the other parameters.

\section{Detailed comparisons of cross power spectra with halo model and TNG}\label{apdx:comp_TNG}

\begin{figure*}
 \includegraphics[width=2\columnwidth]{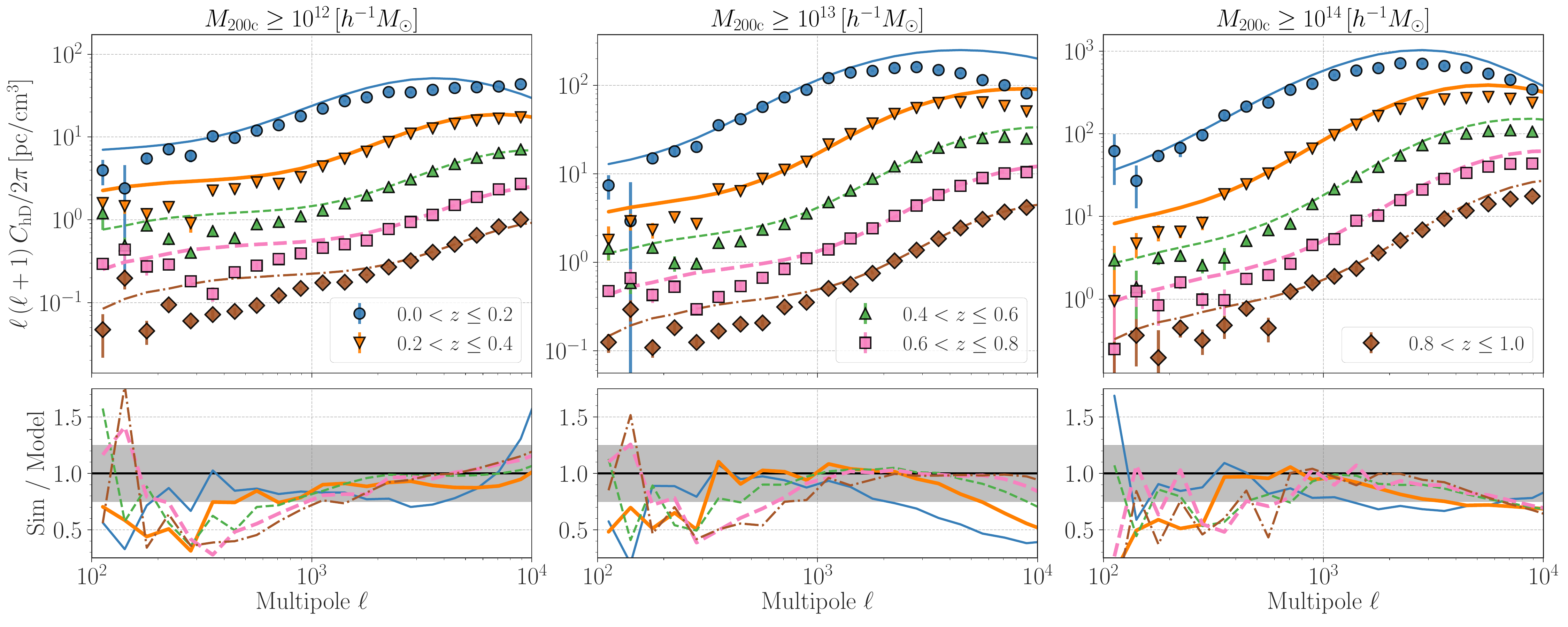}
 \caption{Comparison of the cross power spectra between haloes and DM in TNG simulations with our halo model.
 In each top panel, the different symbols show the power spectra at five redshift bins of $0<z\le0.2$, $0.2<z\le0.4$, $0.4<z\le0.6$, $0.6<z\le0.8$ and $0.8<z\le1.0$; the error bar is the standard deviation for the average over 27 realisations.
 The lines are our model predictions.
 From left to right, we show the comparison for haloes with their masses greater than $10^{12}$, $10^{13}$ and $10^{14}\, h^{-1}M_\odot$.
 To avoid any confusion, we multiply the power spectra at each redshift bin
 by $10^{0}$, $10^{-0.4}$, $10^{-0.8}$, $10^{-1.2}$ and $10^{-1.6}$ from the lowest to highest redshift bins.
 In each bottom panel, we show the ratio of the power spectrum between the simulation results and halo-model counterparts. The grey shaded region in the bottom highlights $\pm25\%$-level differences. 
 }
 \label{fig:TNG_comp_cls_allz}
\end{figure*}

We \en{now} compare our model of the cross power spectrum between dark matter haloes and \en{DMs} with the counterpart in the TNG simulations.

When computing the cross power spectrum, we use 27 realisations of the DM maps at source redshifts being 1 in \citet{2021MNRAS.502.2615T}. 
We also consider three mass-limited samples of haloes for our cross-correlation analysis. We work with the halo samples with their masses greater than $10^{12}$, $10^{13}$ and $10^{14}\, h^{-1}M_\odot$.
We divide the halo redshifts into five bins of $0<z\le0.2$, $0.2<z\le0.4$, $0.4<z\le0.6$, $0.6<z\le0.8$ and $0.8<z\le1.0$ to study the redshift dependence.

For a given halo sample, we construct the over density field in the halo number density 
on $5400\times5400$ grids in a sky of $6\times6 \, {\rm deg}^2$.
Note that the same angular grid is adopted in the DM maps.
For a given set of DM map $D(\bmtheta)$ and 
halo over-density field $\delta^\mathrm{2D}_\mathrm{h}(\bmtheta)$,
we estimate the cross power spectrum as \en{follows:}
\beq
\hat{C}_\mathrm{hD}(\ell) = \frac{1}{N_\ell}\sum_{\bmell;|\bmell|\in\ell} \mathrm{Re}\left[\tilde{\delta}^\mathrm{2D}_\mathrm{h}(\bmell) \tilde{D}^{*}(\bmell)\right], \label{eq:Cest}
\eeq
where the summation runs over modes whose lengths lie in the range
$\ell-\Delta \ell/2 \le |\bmell| \le \ell+\Delta \ell/2$
for the assumed bin width $\Delta \ell$,
$\tilde{\delta}^\mathrm{2D}_\mathrm{h}(\bmell)$ is the Fourier counterpart of $\delta^\mathrm{2D}_\mathrm{h}(\bmtheta)$,
$\tilde{D}^{*}(\bmell)$ is the complex conjugate of $\tilde{D}(\bmell)$
and $\mathrm{Re}[X]$ takes the real part of a given complex number $X$.
In Eq.~(\ref{eq:Cest}), $N_\ell$ represents the number of Fourier modes in a given bin of $\ell$.
We set the width to $\Delta \ln \ell = 0.23$ in the range of $100 < \ell < 10^{4}$.

Figure~\ref{fig:TNG_comp_cls_allz} summarises the comparison of our model of $C_\mathrm{hD}$ and the simulation results.
For the simulation results, we show the average power spectrum over 27 realisations; the error bars represent the standard deviation divided by $\sqrt{27}$.
For a visualisation purpose, we multiply the power spectra \en{in the range of} $0<z\le0.2$, $0.2<z\le0.4$, $0.4<z\le0.6$, $0.6<z\le0.8$ and $0.8<z\le1.0$ by $10^{0}$, $10^{-0.4}$, $10^{-0.8}$, $10^{-1.2}$ and $10^{-1.6}$, respectively.
The figure illustrates that our model can provide a fit to the simulation results in a wide range of $\ell$, halo masses and redshifts.
The model precision reaches about 25\%; overall shapes of the power spectrum are well explained by our halo model.
Our model relies on several fitting formulas of large-scale structures.
We also summarised some action items to develop a more precise model of $C_\mathrm{hD}$ in \en{Subsection}~\ref{subsec:model_accuracy}.

\bibliographystyle{mnras}
\bibliography{refs} 


\bsp	
\label{lastpage}
\end{document}